\documentclass[pra,twocolumn,superscriptaddress]{revtex4-1}
\usepackage{makeidx}
\usepackage[utf8]{inputenc}
\usepackage{graphicx}
\usepackage{amsmath}
\usepackage{mathrsfs}
\usepackage{graphicx}
\usepackage{subfigure}
\usepackage{dcolumn}
\usepackage{bm}
\usepackage{bbm}
\usepackage{color}
\usepackage[dvipsnames]{xcolor}
\usepackage{esint}
\usepackage{lineno}
\usepackage[breaklinks,
            colorlinks,
            urlcolor=blue,
            linkcolor=blue,
            anchorcolor=blue,
            citecolor=blue]{hyperref}

\usepackage{soul}

\begin{document}

\title{Control of spontaneous emission of qubits from weak to strong coupling}
\author{Wai-Keong Mok}
\email{waikeong\_mok@u.nus.edu}
\affiliation{Department of Electronics and Photonics, Institute of High Performance Computing, 1 Fusionopolis Way, 16-16 Connexis,
Singapore 138632, Singapore}
\author{Jia-Bin You}
\email{you\_jiabin@ihpc.a-star.edu.sg}
\affiliation{Department of Electronics and Photonics, Institute of High Performance Computing, 1 Fusionopolis Way, 16-16 Connexis,
Singapore 138632, Singapore}
\author{Wenzu Zhang}
\email{zhangwz@ihpc.a-star.edu.sg}
\affiliation{Department of Electronics and Photonics, Institute of High Performance Computing, 1 Fusionopolis Way, 16-16 Connexis,
Singapore 138632, Singapore}
\author{Wan-Li Yang}
\affiliation{State Key Laboratory of Magnetic Resonance and Atomic and Molecular Physics, Wuhan Institute of Physics and Mathematics, Chinese
Academy of Sciences, Wuhan 430071, China}
\author{Ching Eng Png}
\affiliation{Department of Electronics and Photonics, Institute of High Performance Computing, 1 Fusionopolis Way, 16-16 Connexis,
Singapore 138632, Singapore}

\begin{abstract}

Photon emission and absorption by an individual qubit are essential elements for the quantum manipulation of light. Here we demonstrate the controllability of spontaneous emission of a qubit in various electromagnetic environments. The parameter regimes that allow for flexible control of the qubit emission routes are comprehensively discussed. By properly tuning the system couplings and decay rates, the spontaneous emission rate of the qubit can undergo Purcell enhancement and inhibition. Particularly, when the cavity is prepared in the excited state, the spontaneous emission rate of the qubit can be significantly suppressed. We also demonstrate a spectral filter effect which can be realised by controlling the steady-state emission spectra of qubits.

\end{abstract}

\maketitle

\section{Introduction}

Atom-photon quantum interfaces are key building blocks for a quantum network, with applications including quantum sensing \cite{Maccone:2011aa}, cryptography \cite{al:2017aa} and quantum information processing \cite{Kimble:2008aa}. The basic component of a quantum network comprises local nodes (atoms) used to store or process quantum information. These nodes are connected by quantum channels for photons (``flying qubits") to be used to distribute quantum information \cite{Kimble:2008aa,J.I.-Cirac:1997aa,S.-Ritter:2012aa,Duan:2010aa} and entanglement \cite{E.-Knill:2001aa} over the network. Information transfer can be accomplished by either directly absorbing the flying qubit \cite{J.I.-Cirac:1997aa} or through projective measurements \cite{C.-Cabrillo:1999aa,M.-B.-Plenio:1999aa,Simon:2003aa,X.-L.-Feng:2003aa}. As a result, controlling fundamental processes of photon emission and absorption by qubits \cite{C.-Cohen-Tannoudji:2004aa} is essential to utilise quantum information.

Over the years, many quantum optical schemes have been developed for the temporal and spectral manipulations of the photon emission and absorption processes of qubits. \cite{Keller2004,Darquie454,PhysRevLett.103.213601,PhysRevA.96.043807,PhysRevA.90.023829,1367-2630-15-5-053007,1367-2630-15-5-055005}. Among these systems, cavity and circuit quantum electrodynamic systems \cite{Carmichael:1993aa,Putra:2005aa} are promising candidates for distributed quantum information processing \cite{J.-I.-Cirac:1999aa,A.-Reiserer:2014aa} due to their advantages of \textit{in situ} tunability and individual addressing of cavity/circuit elements \cite{RevModPhys.84.1}, spectroscopic technology for state readout \cite{Jiang267}, and scalability \cite{Houck2012,PhysRevA.86.023837}. In recent years, these systems displayed a wide range of experimental results,
from photon blockade and single-photon generation to quantum state transfer between atoms and photons \cite{S.-Ritter:2012aa,Hamsen:2017aa,A.-Reiserer:2013aa,A.-Reiserer:2014aa,Parkins:2014aa}; from long distance quantum networks \cite{S.-Welte:2018aa} to scalable quantum network which is essential for applications in quantum computing \cite{Raussendorf:2001aa,H.P.-Buchler:2005aa}. Strong nonlinearity of the system, which is an essential requirement for these applications, can be achieved in the strong coupling regime \cite{Mabuchi:2002aa,R.-Miller:2005aa}.

In this paper, we comprehensively study the manipulation of the spontaneous emission (SE) of qubits in various electromagnetic environments, including another qubit and cavity, from weak to strong coupling regime. The equation-of-motion method is applied to obtain various propagators for single and double excitations of the system. For the spontaneous emission, the Purcell enhancement and inhibition are observed in the temporal domain; particularly, we find that preparing the electromagnetic environment in the excited state will significantly slow down the spontaneous emission of the qubit. Furthermore, examining each involved propagator provides more insights on the emission routes of the qubit. We also give some frequency domain results demonstrating how the emission spectra can be utilised as a frequency filter when the system reaches steady state. The physical insights developed here allow for great control over the photon emission process of the qubit which is useful in the fields of quantum information and communication.

The paper is organised as follows. In Section \ref{sec:model} we introduce the general model for describing two qubits in a cavity. In the subsequent sections we focus on the dynamical evolution of transition probabilities for several special cases which are derived by simplifying the original model. This includes the two-qubit model in Section \ref{sec:two_qubits} and the two-excitation Jaynes-Cummings model (JCM) in Section \ref{sec:JCM_n=1}. The general model is treated in detail in Section \ref{sect:eg+cavity}, which includes a discussion of our steady-state results in the spectral domain. We conclude the results in Section \ref{sec:conclusion}, and the appendices provide further details on the calculations made in above sections.

\section{Model and Hamiltonian}
\label{sec:model}

The system in consideration is basically the Tavis-Cummings model (TCM) \cite{PhysRev.170.379,Chen:2009aa} where only two qubits (labelled as 1 and 2) are coupled to a single mode cavity. The qubits couple via dipole-dipole interactions $g_{12}$ and interact with the cavity via couplings $g_{1}$ and $g_{2}$ respectively. The Hamiltonian is given by (setting $\hbar = 1$)
\begin{equation}
\label{H0}
\begin{split}
H_{\text{s}}&=-\tfrac{\omega_{01}}{2}|{g_1}\rangle\langle{g_1}|-\tfrac{\omega_{02}}{2}|{g_2}\rangle\langle{g_2}|+g_{12}|{g_{1}e_{2}}\rangle\langle{e_{1}g_{2}}|\\
&+\tfrac{\omega_{c}}{2}a^\dag{a}+g_1a^\dag|{g_1}\rangle\langle{e_1}|+g_{2}a^\dag|{g_2}\rangle\langle{e_2}|+\text{H.c.},\\
\end{split}
\end{equation}	
where the transition frequencies of the qubits and the resonant frequency of the cavity are $\omega_{01}$, $\omega_{02}$ and $\omega_{c}$ respectively. The energies of the atomic excited states are defined as identically zero. The bosonic operators for the cavity mode are $a^\dag$ and $a$, satisfying the canonical commutation relation $[a, a^\dag] = 1$.

To study the open system dynamics of the system, the qubits and cavity are coupled by three independent baths,
\begin{equation}
\label{H1}
\begin{split}
H_{\text{b}}&=\sum_{i=1}^{3}\int_0^\infty\mathrm{d}\omega \, \omega b_i^\dag(\omega)b_i(\omega),\\
\end{split}
\end{equation}
where $b_i^\dag (\omega), b_i (\omega), (i = 1,2,3)$ are bath operators following the canonical commutation relation $[b_i (\omega), b_{i^\prime}^\dag (\omega^\prime)] = \delta_{i i^\prime} \delta(\omega - \omega^\prime)$. In this work, we consider the spontaneous emission of the system to the surrounding vacuum which can be modeled as an independent bath since the vacuum is an uncorrelated environment. The couplings between bath and system are
\begin{equation}
\label{H01}
\begin{split}
H_{\text{s-b}}&=\int_0^\infty \mathrm{d}\omega \, \left[\sqrt{\frac{\Gamma_1}{2\pi}} |{e_1}\rangle\langle{g_1}| b_1(\omega) + \sqrt{\frac{\Gamma_2}{2\pi}}\,|{e_2}\rangle\langle{g_2}| b_2(\omega)\right] \\
&+ \sqrt{\frac{\kappa}{2\pi}} \int_0^\infty \mathrm{d}\omega \, a b_3^\dag (\omega)+ \text{H.c.}, \\
\end{split}
\end{equation}

where $\Gamma_{1}$, $\Gamma_{2}$ and $\kappa$ are the decay rates of qubits and cavity in the Wigner-Weisskopf approximation \cite{Muller:2017aa}. Thus the total Hamiltonian for the open system is given by $H=H_{\text{s}}+H_{\text{b}}+H_{\text{s-b}}$.

In the following, we shall study the manipulation of the spontaneous emission of the initially excited qubit 1 in various electromagnetic environments (EME) including i) another qubit; ii) a cavity and iii) Jaynes-Cummings environment where each EME is prepared in its ground state and/or single-excited state.

\section{Two-qubit system}
\label{sec:two_qubits}

\subsection{Qubit 2 prepared in ground state}
\label{two_qubits_single_excitation}

\begin{figure}
\includegraphics[width=6cm]{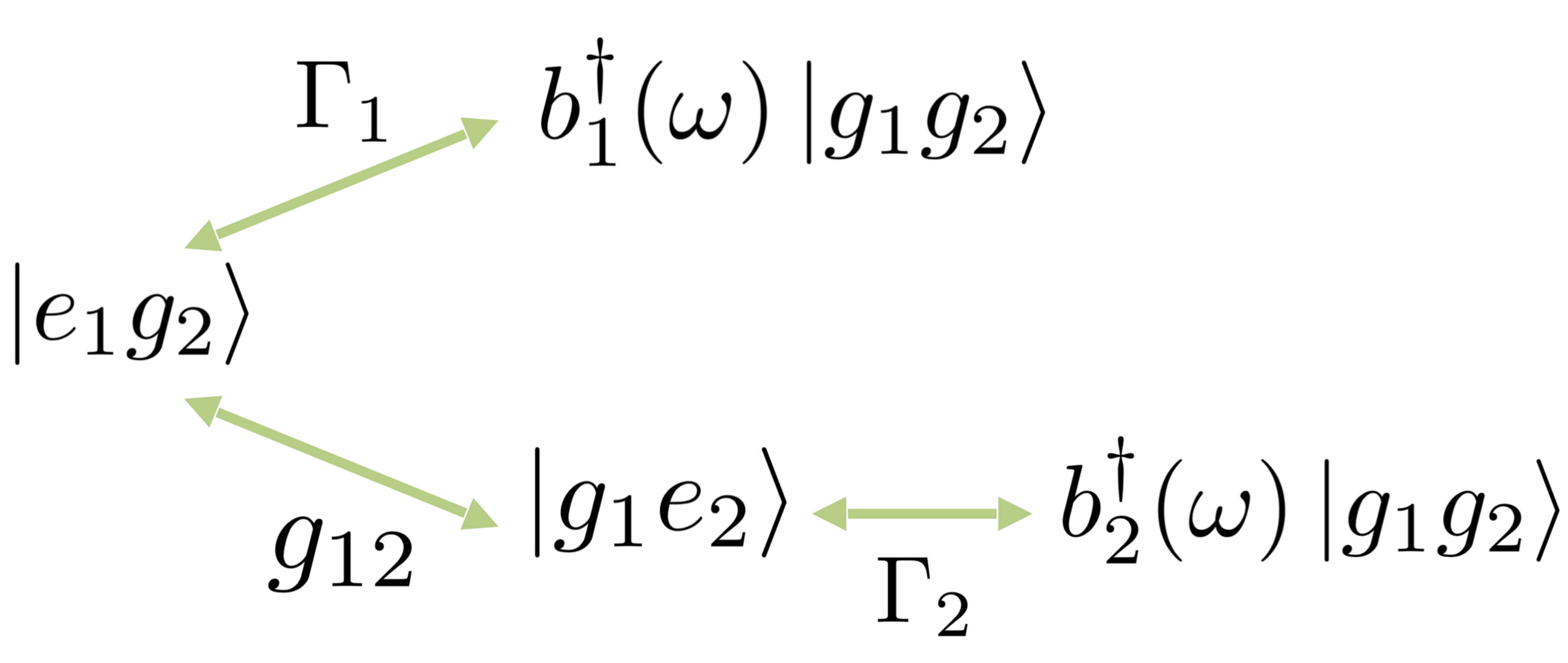}
\caption{(color online). Network of quantum states for spontaneous emission of qubit 1.}
\label{fig:eg_network}
\end{figure}

Decoupling the cavity from the qubits, the system simplifies to two coupled qubits without cavity. To manipulate the spontaneous emission of qubit 1, we prepare qubit 1 in the excited state $|e_{1}\rangle$ and study the decay probabilities via different channels by the equation-of-motion method. We first prepare qubit 2 initially in the ground state $|g_{2}\rangle$. Thus the initial state of both qubits is $|e_{1} g_{2}\rangle$. Using the Schr\"{o}dinger equation $i \frac{\partial}{\partial t} U(t) = H U(t)$, where $U(t) =e^{-iHt}$ is the propagator, the resulting system of differential equations for SE of qubit 1 with final state $b_1^\dag (\omega_1) |{g_{1}g_{2}}\rangle$ reads
\begin{widetext}
\begin{equation}
\label{EOM}
\begin{split}
&i \frac{\partial}{\partial t} \langle{g_1 g_2}|  b_1 (\omega_1) U(t) |{e_1 g_2}\rangle = -\omega_{02} \langle{g_1 g_2}|  b_1 (\omega_1) U(t) |{e_1 g_2}\rangle + \sqrt{\frac{\Gamma_{1}}{2\pi}} \int_0^\infty \mathrm{d} \omega \langle{g_1 g_2}| b_1 (\omega_1) U(t) b_1^\dag (\omega) |{g_1 g_2}\rangle \\
&+ g _{12}\langle{g_1 g_2}| b_1 (\omega) U(t) |{g_1 e_2}\rangle, \\
&i \frac{\partial}{\partial t} \langle{g_1 g_2}| b_1 (\omega_1) U(t) b_1^\dag (\omega) |{g_1 g_2}\rangle = (\omega - \omega_{01} - \omega_{02}) \langle{g_1 g_2}| b_1 (\omega_1) U(t) b_1^\dag (\omega) |{g_1 g_2}\rangle + \sqrt{\frac{\Gamma_{1}}{2\pi}} \langle{g_1 g_2}| b_1 (\omega_1) U(t) |{e_1 g_2}\rangle, \\
&i \frac{\partial}{\partial t} \langle{g_1 g_2}| b_1(\omega_1) U(t) |{g_1 e_2}\rangle = -\omega_{02} \langle{g_1 g_2}| b_1 (\omega_1) U(t) |{g_1 e_2}\rangle + g_{12} \langle{g_1 g_2}| b_1(\omega_1) U(t) |{e_1 g_2}\rangle \\
&+ \sqrt{\frac{\Gamma_{2}}{2\pi}} \int_0^\infty \mathrm{d}\omega \langle{g_1 g_2}| b_1(\omega_1) U(t) b_2^\dag(\omega) |{g_1 g_2}\rangle, \\
&i \frac{\partial}{\partial t} \langle{g_1 g_2}| b_1(\omega_1) U(t) b_2^\dag(\omega) |{g_1 g_2}\rangle = (\omega - \omega_{01} - \omega_{02}) \langle{g_1 g_2}| b_1 (\omega_1) U(t) b_2^\dag (\omega) |{g_1 g_2}\rangle + \sqrt{\frac{\Gamma_{2}}{2\pi}} \langle{g_1 g_2}| b_1 (\omega_1) U(t) |{g_1 e_2}\rangle. \\
\end{split}
\end{equation}
\end{widetext}
Note that the propagators involved in the SE can be organized as a network as depicted in Fig. \ref{fig:eg_network}. It is shown that the quantum states are connected by either a decay or coupling process.
Using the Laplace transform to convert the differential equations in Eq. (\ref{EOM}) into algebraic equations, the spontaneous emission probability $P_{\text{SE}} (t, \omega_1) = | \langle{g_{1}g_{2}}| b_1 (\omega_1) U(t) |{e_{1}g_{2}}\rangle |^2$ is readily solved as
\begin{equation}
\begin{split}
P_{\text{SE}} (t, \omega_1) &= \frac{\Gamma_{1}}{2\pi} \bigg| \frac{ \omega_1 + i\frac{\Gamma_2}{2} - \omega_{02}}{ (\Delta_1 - \omega_+)(\Delta_1 - \omega_-)} e^{-i \Delta_1 t } \\
&+ \frac{ \omega_+ + i\frac{\Gamma_2}{2} + \omega_{01}}{(\omega_+ - \Delta_1)(\omega_+ - \omega_-)} e^{-i \omega_+ t} \\
&- \frac{\omega_- + i\frac{\Gamma_2}{2} + \omega_{01}}{(\omega_- - \Delta_1)(\omega_+ - \omega_-)} e^{-i \omega_- t} \bigg|^2,\\
\end{split}
\label{eqn:JC_n=0_SE}
\end{equation}
where $\Delta_1 \equiv \omega_1 - \omega_{01} - \omega_{02}$ and
\begin{equation}
\begin{split}
\omega_\pm &= \frac{1}{2} \bigg[ -\bigg(i\frac{\Gamma_1}{2} + i\frac{\Gamma_2}{2} + \omega_{01} + \omega_{02}\bigg) \\
&\pm \sqrt{ \bigg( i\frac{\Gamma_1}{2} - i\frac{\Gamma_2}{2} + \omega_{02} - \omega_{01}\bigg)^2 + 4g_{12}^2} \bigg]
\end{split}
\end{equation}
are the dressed state eigenfrequencies. We can also obtain the total spontaneous emission probability of qubit 1 at a particular time $P_{\text{SE}}(t)$ by integrating $P_{\text{SE}} (t, \omega_1)$ over all emission frequencies, i.e. $P_{\text{SE}}(t)=\int_{-\infty}^\infty \mathrm{d} \omega_1 P_{\text{SE}} (t, \omega_1)$. Similarly, all the other probabilities involved in the SE can be derived,
\begin{equation}
\label{probs_in_SE}
\begin{split}
&P_{\text{surv}} (t) = |\langle{e_{1}g_{2}}|U(t)|{e_{1}g_{2}}\rangle|^2 \\
&= \left|\frac{\omega_+ + i\frac{\Gamma_2}{2} + \omega_{01}}{\omega_+ - \omega_-} e^{-i\omega_+ t} - \frac{\omega_- + i\frac{\Gamma_2}{2} + \omega_{01}}{\omega_+ - \omega_-} e^{-i\omega_- t}\right|^2,\\
&P_{\text{exchg}} (t) = |\langle{g_{1}e_{2}}| U(t) |{e_{1}g_{2}}\rangle|^2 \\
&= \bigg|\frac{g_{12}}{\omega_+ - \omega_-} \left( e^{-i\omega_+ t} - e^{-i \omega_- t} \right) \bigg|^2,\\
&P_{\text{em2}} (t) = \int_{-\infty}^\infty \mathrm{d} \omega_r \, |\langle{g_{1}g_{2}}| b_2 (\omega_r) U(t) |{e_{1}g_{2}}\rangle|^2 \\
 &= g_{12}^2 \frac{\Gamma_{2}}{2\pi} \int_{-\infty}^\infty \mathrm{d} \omega_r \, \bigg| \frac{1}{(\omega_+ - \omega_-)(\omega_+ - \Delta_2)} e^{-i\omega_+ t} \\
 &+ \frac{1}{(\omega_+ - \omega_-)(\omega_- - \Delta_2)} e^{-i \omega_- t} \\
 &+ \frac{1}{(\Delta_2 - \omega_+)(\Delta_2 - \omega_- )} e^{-i\Delta_2 t} \bigg|^2,
\end{split}
\end{equation}
which are the probabilities for excited state survival of qubit 1, excitation exchange between the qubits, and exchange-emission mediated by qubit 2, respectively. Here $\Delta_2 \equiv \omega_2 - \omega_{01} - \omega_{02}$. Now we shall discuss the above probabilities in detail.

\begin{figure}
\subfigure{\includegraphics[width=4cm]{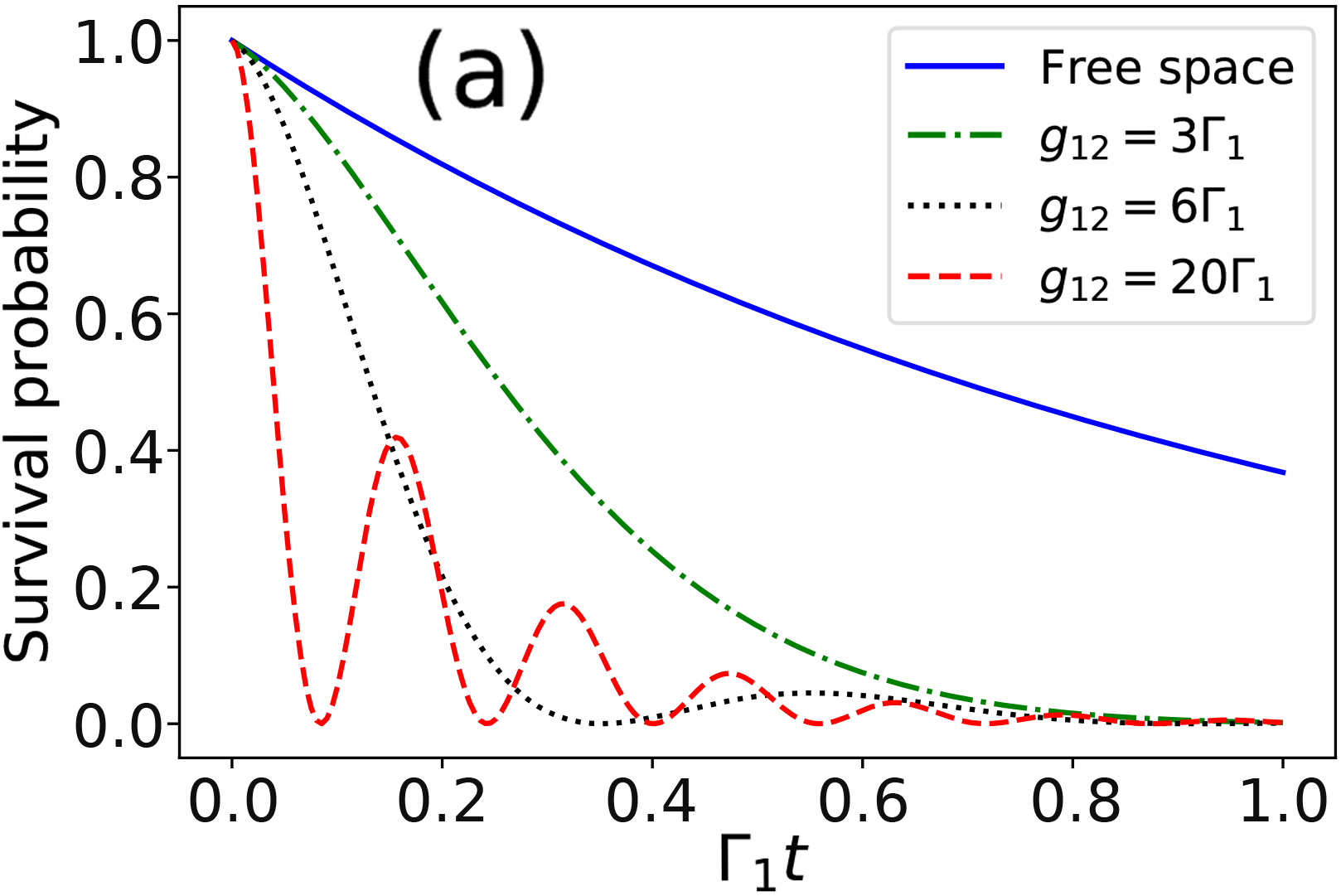}
\label{fig:purcell_rabi_time(a)}}
\subfigure{\includegraphics[width=4cm]{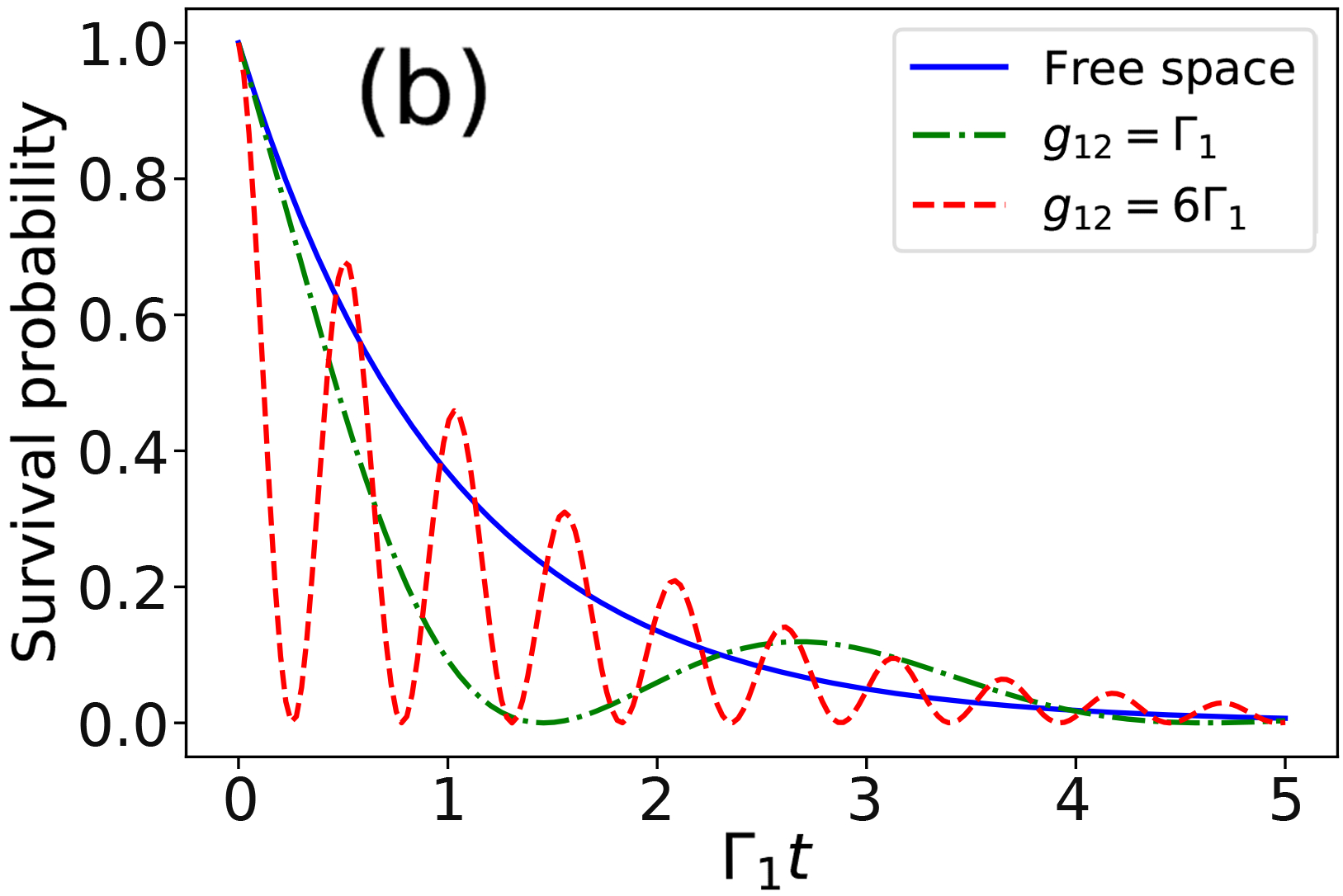}
\label{fig:purcell_rabi_time(b)}} \\
\subfigure{\includegraphics[width=4cm]{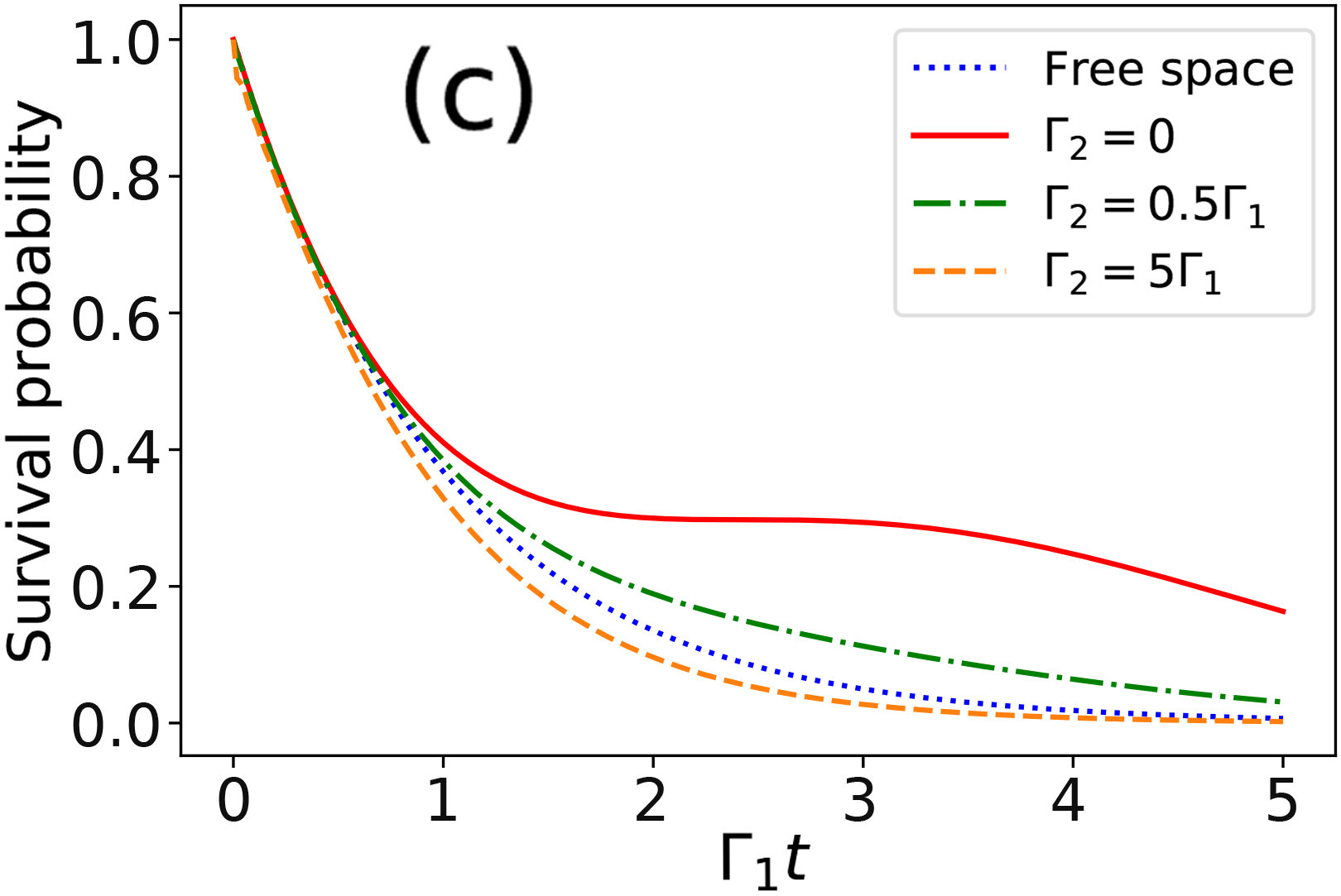}
\label{fig:purcell_rabi_time(c)}}
\subfigure{\includegraphics[width=4cm]{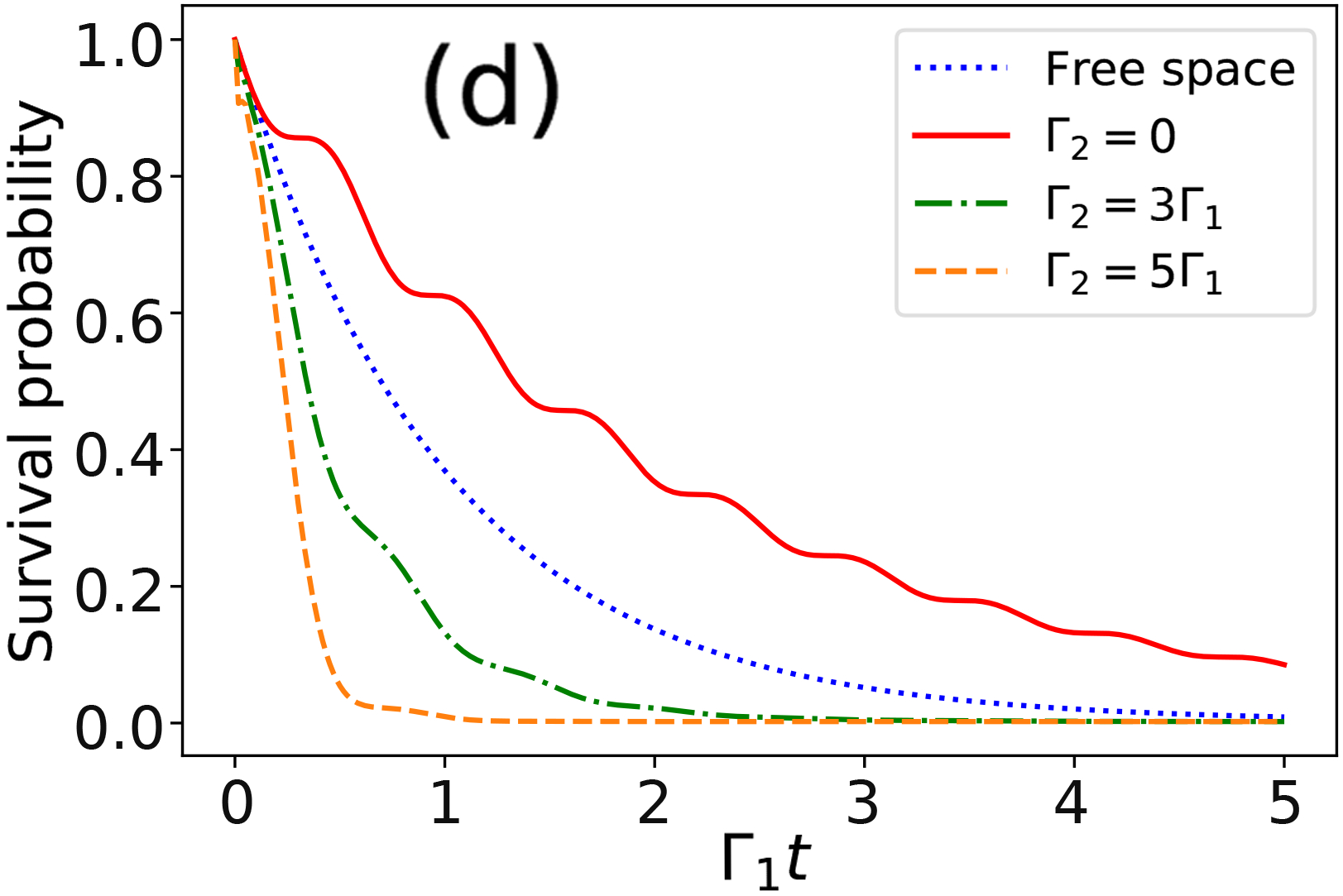}
\label{fig:purcell_rabi_time(d)}} \\
\subfigure{\includegraphics[width=4cm]{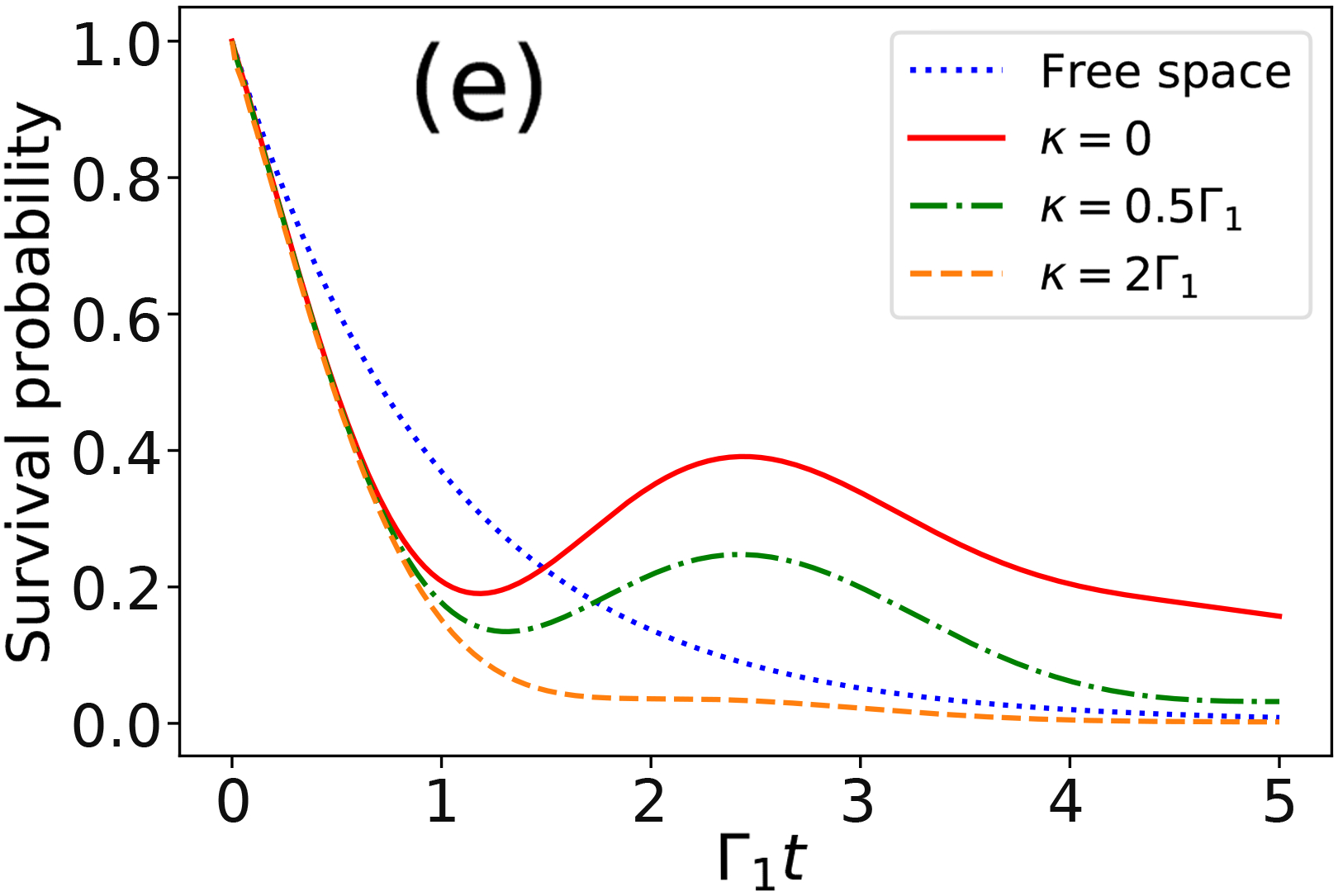}
\label{fig:purcell_rabi_time(e)}}
\subfigure{\includegraphics[width=4cm]{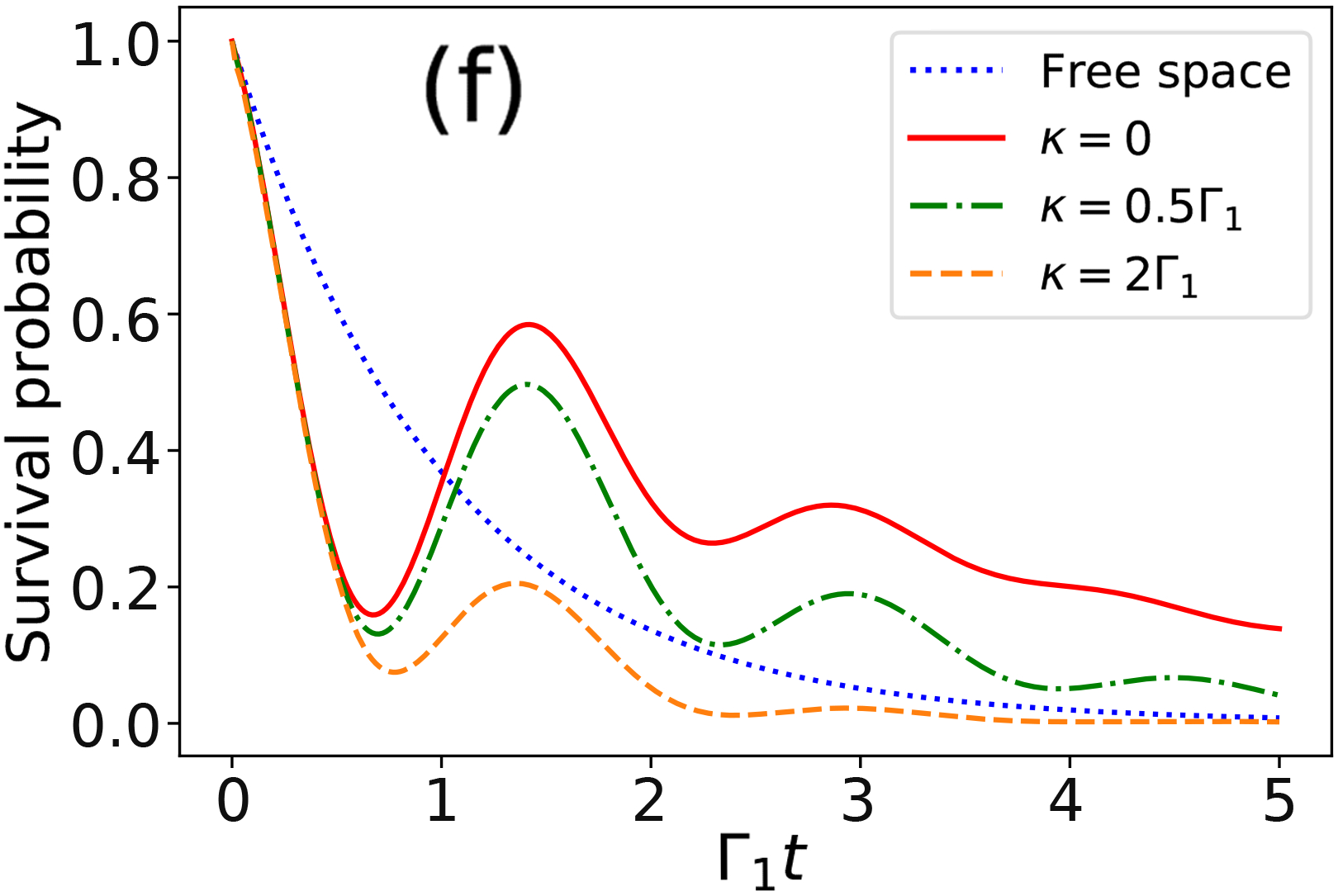}
\label{fig:purcell_rabi_time(f)}}\\
\subfigure{\includegraphics[width=4cm]{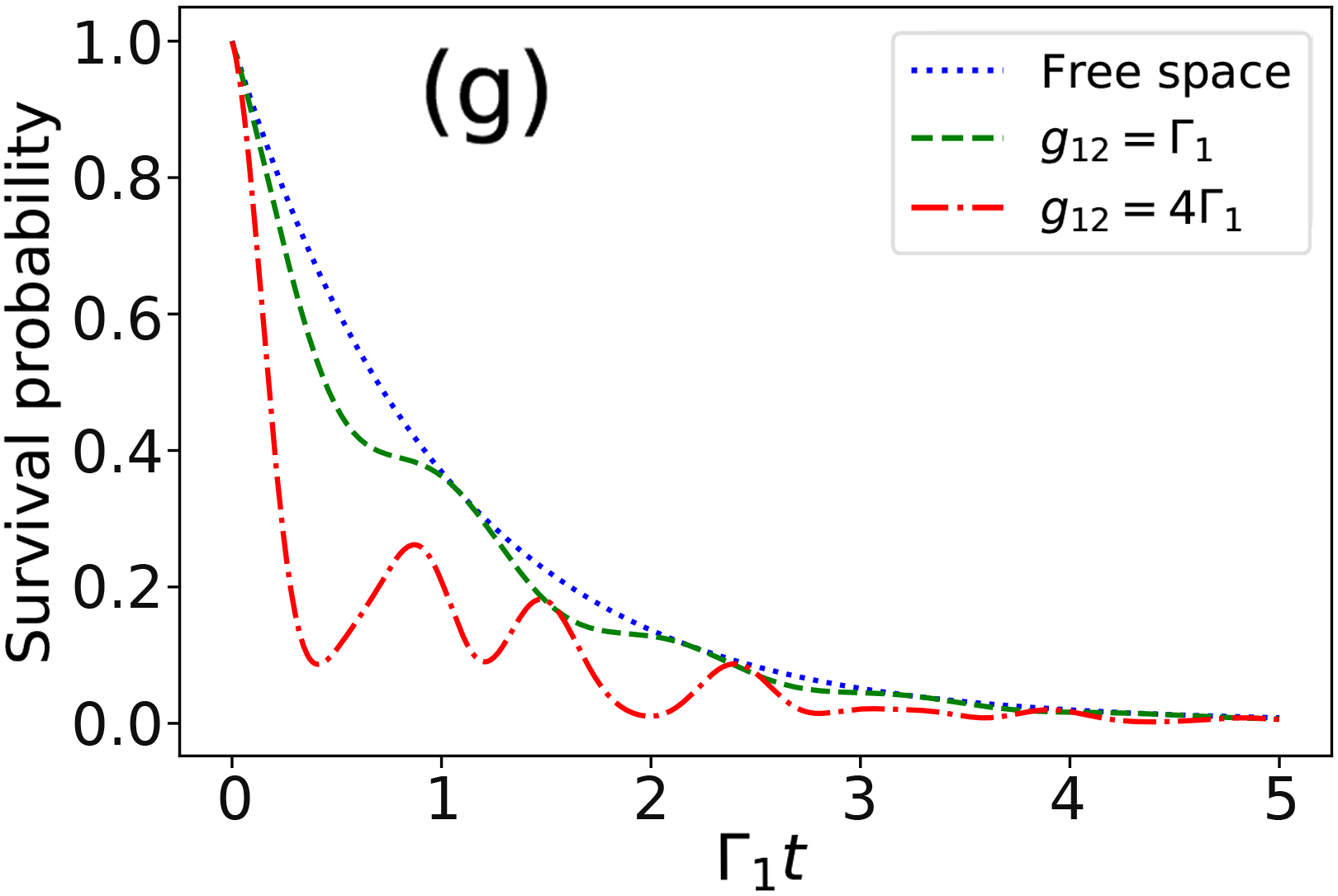}
\label{fig:purcell_rabi_time(g)}}
\subfigure{\includegraphics[width=4cm]{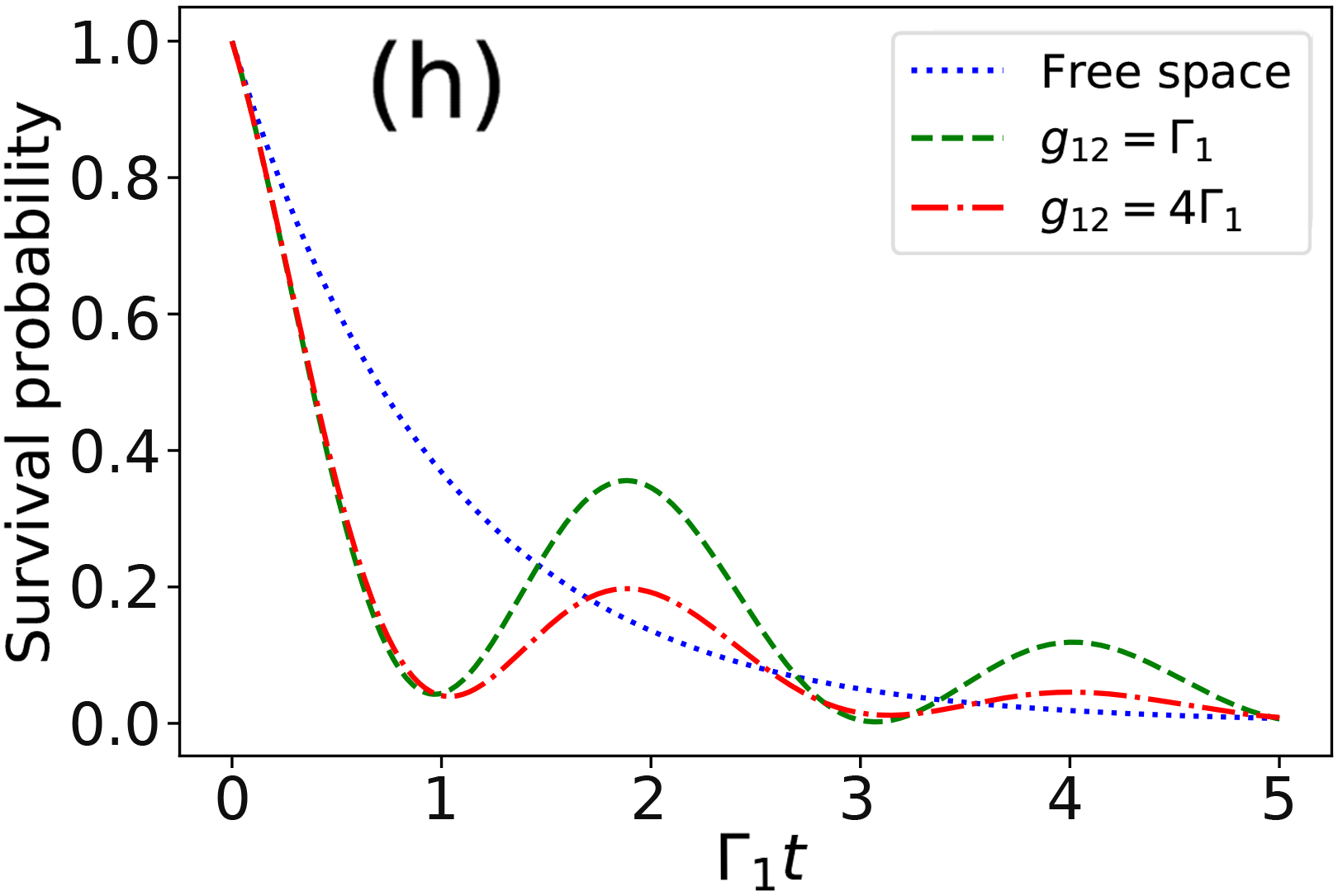}
\label{fig:purcell_rabi_time(h)}} \\
\caption{(color online). Time evolution of survival probability $P_{\text{surv}} (t)$ of qubit 1 with resonant qubits ($\omega_{01} = \omega_{02}$). (a) and (b) correspond to the two-qubit, single-excitation system (Section \ref{two_qubits_single_excitation}). (c) and (d) correspond to the two-qubit, two-excitations system (Section \ref{two_qubit_double_excitations}). (e) and (f) correspond to the two-excitation JCM (Section \ref{sec:JCM_n=1}). (g) and (h) correspond to the TCM (Section \ref{sect:eg+cavity}). Parameters used: (a) $\Gamma_2 = 10 \Gamma_1$, bad atom regime. (b) $\Gamma_2 = 0.5 \Gamma_1$, good atom regime. (c) $g_{12} = 0.5\Gamma_1$, weak coupling regime, (d) $g_{12} = 5\Gamma_1$, strong coupling regime. (e) $g = \Gamma$, weak coupling regime, (f) $g = 2\Gamma$, strong coupling regime. (g) $g_1 = \Gamma_1, g_2 = 5\Gamma_1, \kappa = \Gamma_1, \Gamma_2 = \Gamma_1$, (h) $g_1 = \Gamma_1, g_2 = \Gamma_1, g_{12} = \Gamma_1, \kappa = 0$.}
\label{fig:purcell_rabi_time}
\end{figure}

First, we demonstrate that by tuning the coupling strength and decay rates, the spontaneous emission rate can be efficiently modified. The plots in Fig. \ref{fig:purcell_rabi_time} are for the analytical solutions from the Laplace transform method, which are checked to agree with the numerical solutions to the Schr\"{o}dinger equation. In Fig. \ref{fig:purcell_rabi_time}(a) where qubit 1 is resonantly coupled with qubit 2, the survival probability of qubit 1 $P_{\text{surv}}(t)$ decays faster compared to the free-space emission when $\Gamma_1 < \Gamma_2$. This is the well-known Purcell enhancement \cite{Vahala:2003aa}. In the strong coupling regime where $g_{12} > (\Gamma_1, \Gamma_2)$, it is expected that vacuum Rabi oscillations are observed.
In contrast to Purcell enhancement, the inhibition of spontaneous emission of qubit 1 can be observed when $\Gamma_1 > \Gamma_2$. Due to Rabi oscillations, the survival probability exceeds the free-space value at certain times, as shown in Fig. \ref{fig:purcell_rabi_time(b)}. The mechanism is that when the decay rate of qubit 2 is smaller than the decay rate of qubit 1, the capacity of energy storage for qubit 2 is better than that for qubit 1. Qubit 2 serves as an energy storage to repeatedly feedback the energy to qubit 1 via $g_{12}$ coupling which suppresses spontaneous emission of qubit 1. Thus, we can control the SE of qubit 1 by tuning the decay rate of qubit 2 in the strong coupling regime.

\begin{figure*}
\centering
\subfigure{\includegraphics[width=5.3cm]{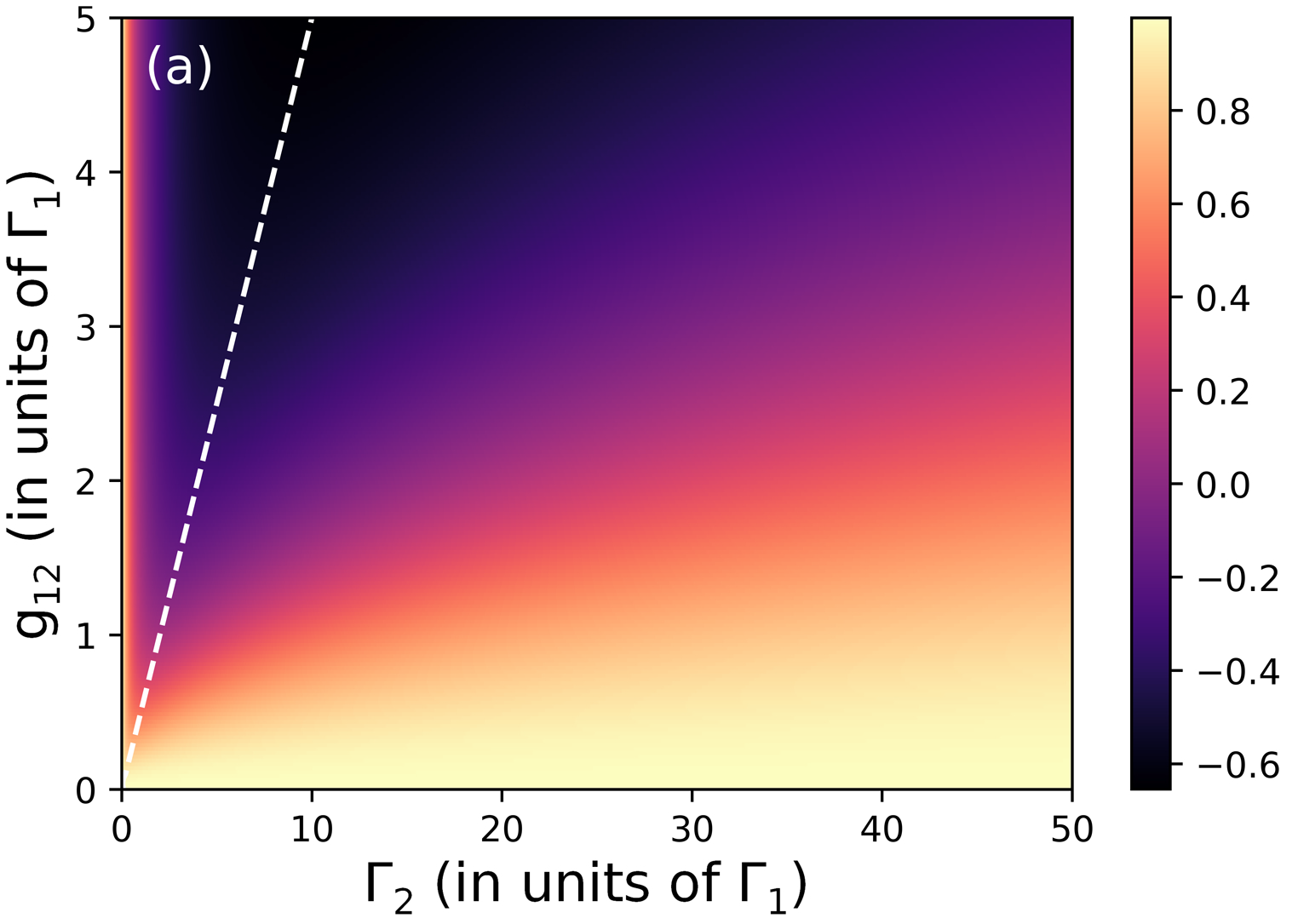}\label{fig:jc_gapmap}}
\subfigure{\includegraphics[width=5.3cm]{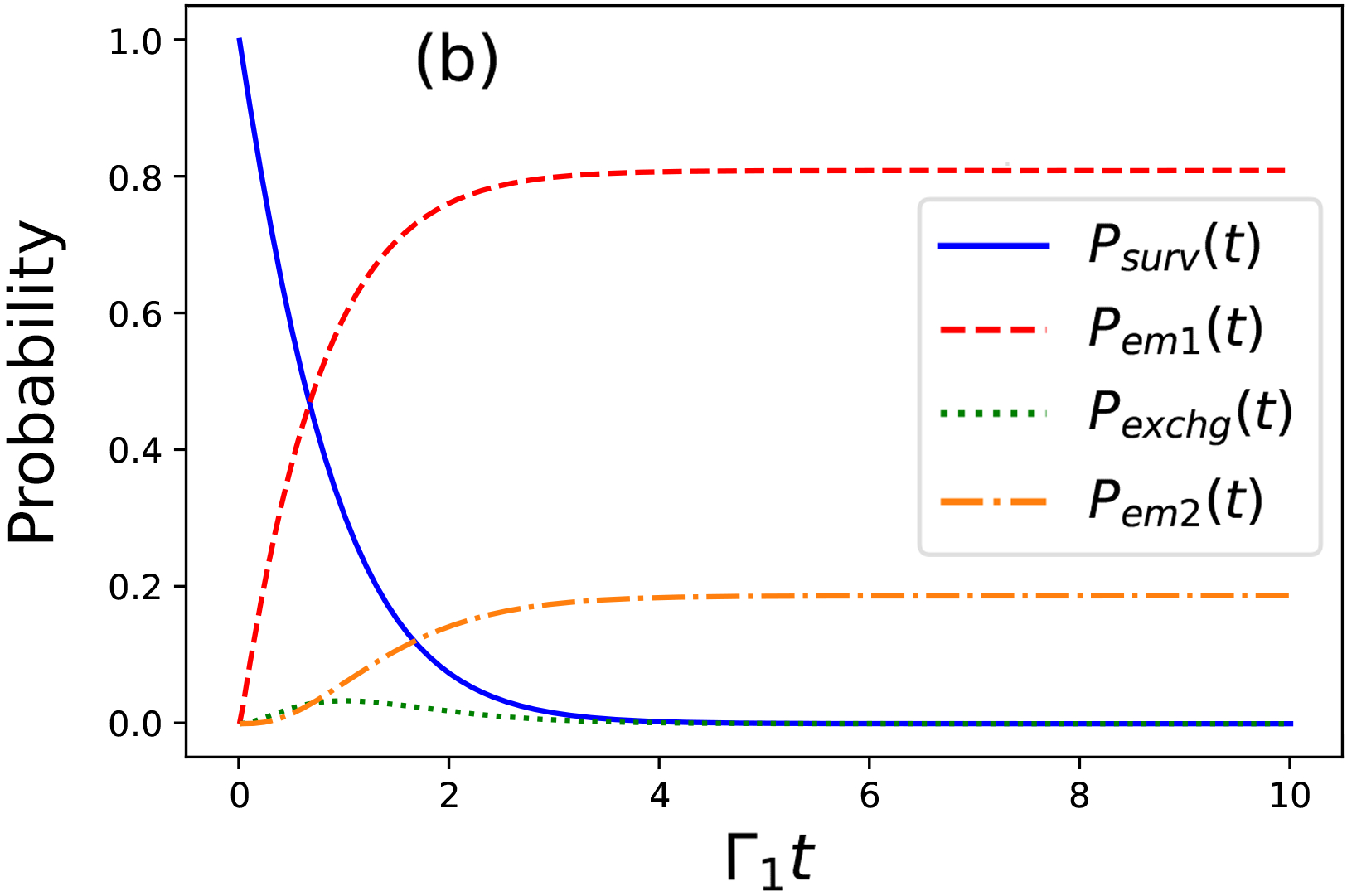}\label{fig:jc_combtime_b}}
\subfigure{\includegraphics[width=5.3cm]{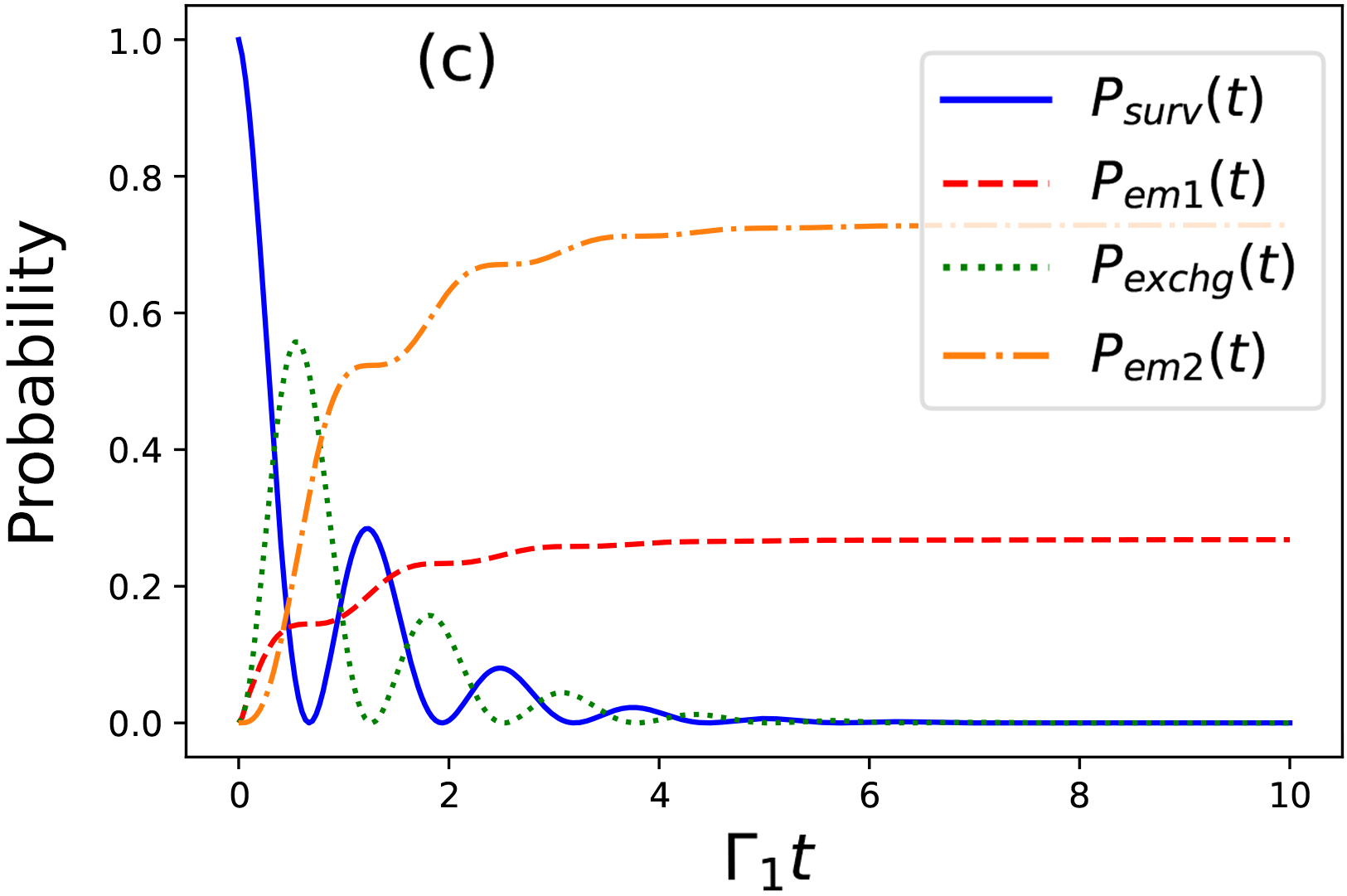}\label{fig:jc_combtime_c}}
\caption{(color online). (a) Variation of the gap $P_{\text{SE}}(t\rightarrow\infty)- P_{\text{em2}}(t\rightarrow\infty)$ between the qubit 1 decay probability and qubit 2 decay probability in steady state as a function of $\Gamma_2$ and $g_{12}$. The white dashed line indicates the optimal $\Gamma_2$ which maximises $P_\text{em2}(t \to \infty)$ for each $g_{12}$. The parameters for time evolution of probabilities $P_{\text{SE}}(t)$ and $P_{\text{em2}}(t)$ in Eq. (\ref{eqn:JC_n=0_SE}) and Eq. (\ref{probs_in_SE}) are (b) $g_{12} = 0.5\Gamma_1, \Gamma_2 = 3\Gamma_1$ (weak coupling regime); (c) $g_{12} = 5\Gamma_1, \Gamma_2 = 3\Gamma_1$ (strong coupling regime).}
\label{fig:jc_combtime}
\end{figure*}

Next, we discuss the dominant emission route during the SE of qubit 1. The system has two emission routes: qubit 1 emission via $\Gamma_1$, and qubit 2 emission via $\Gamma_2$. Fig. \ref{fig:jc_combtime_b} shows the dynamical evolution of the different probabilities in Eq. (\ref{probs_in_SE}) in the weak coupling regime. It is shown that the photon is more likely to emit via qubit 1, even though $\Gamma_2 > \Gamma_1$. In the strong coupling regime however, the decay via qubit 2 dominates as depicted in Fig. \ref{fig:jc_combtime_c}. This shows that the coupling strength is more crucial than $\Gamma_2$ in determining the relative dominance of emission routes of qubit 1. The difference between $P_{\text{SE}}(t)$ and $P_{\text{em2}}(t)$  in the steady state provides a quantitative measure of this behaviour. This probability gap $P_{\text{SE}}(t\rightarrow\infty)- P_{\text{em2}}(t\rightarrow\infty)$ is shown in the contour plot in Fig. \ref{fig:jc_gapmap}. For any value of $g_{12}$, increasing $\Gamma_2$ will first cause the decrease of the gap which indicates that qubit 2 starts to emit photons; the gap increases again when qubit 2 enters the bad-atom regime (large $\Gamma_2$) where Purcell enhancement of qubit 1 is expected. For weak coupling, the gap is always positive, indicating that qubit 1 relaxation dominates over qubit 2 relaxation. Under strong coupling, it is interesting that the probability gap can reach negative values, indicating that the emission route via qubit 2 outweighs that of qubit 1. This phenomenon is exclusive for strong coupling which is imprinted in the oscillation behavior in the time evolution of probabilities as shown in Fig. \ref{fig:jc_combtime_c}. Our results also suggest that for a value of $g_{12}$, there exists an optimal $\Gamma_2$ which maximises the steady-state probability of qubit 2 emission $P_\text{em2} (t \to \infty)$. Note that the normalisation condition requires $P_\text{em2} (t \to \infty) + P_\text{em1} (t \to \infty) = 1$. The variation of the optimal $\Gamma_2$ with $g_{12}$ is indicated by the white dashed line in Fig. \ref{fig:jc_gapmap}.

\subsection{Qubit 2 prepared in excited state}
\label{two_qubit_double_excitations}

The situation becomes more interesting if the EME of qubit 1, i.e., qubit 2 is also prepared in the excited state such that both qubits are initially excited, i.e., state $|{e_{1}e_{2}}\rangle$. In this case, qubit 2, now in a higher energy state, serves as an energy charger to pump energy into qubit 1 and prolong its lifetime.

To analyse the effect of the excited qubit 2 on the spontaneous emission of the qubit 1, we consider the survival probability of qubit 1 in its excited state $|e_{1}\rangle$. Three propagators are involved in this process: the survival probability of state $|{e_{1}e_{2}}\rangle$, $P_{\text{surv}} (t) = | \langle{e_{1}e_{2}}| U(t) |{e_{1}e_{2}}\rangle |^2$; the single-photon emission probability via qubit 2, $P_{\text{em2}} (t, \Delta_2) = | \langle{e_{1}g_{2}}| b_2 (\omega_2) U(t) |{e_{1}e_{2}}\rangle |^2$ and the emission-exchange probability $P_{\text{emx1}} (t, \Delta_1) = | \langle{e_{1}g_{2}}| b_1 (\omega_1) U(t) |{e_{1}e_{2}}\rangle |^2$. The survival probability of qubit 1 is then given by $P_{\text{total}} (t) = P_{\text{surv}} (t) + P_{\text{em2}} (t) + P_{\text{emx1}} (t)$, where $P_{\text{em2}} (t)$ and $P_{\text{emx1}} (t)$ are obtained by integrating their frequency-resolved probabilities over their respectively emission frequencies. By the Schr\"odinger equation as above, the survival probability of state $|{e_{1}e_{2}}\rangle$ reads $P_{\text{surv}} (t)=e^{-(\Gamma_1 + \Gamma_2)t}$.  The frequency-resolved single-photon emission probability and emission-exchange probability read
\begin{equation}
\label{Pem2_2qubits}
\begin{split}
P_{\text{em2}} (t, \Delta_2) &= \frac{\Gamma_2}{2\pi} \bigg| \frac{\omega_{2+} - \omega_2 + \omega_{01} + i\frac{\Gamma_2}{2}}{(\omega_{2+} - \omega_{2-})(\omega_{2+} + i \Gamma)} e^{-i\omega_{2+} t} \\
&+ \frac{\omega_{2-} - \omega_2 + \omega_{01} + i\frac{\Gamma_2}{2}}{(\omega_{2-} - \omega_{2+})(\omega_{2-} + i\Gamma)} e^{-i\omega_{2-} t} \\
&- \frac{i\Gamma + \omega_2 - \omega_{01} - i\frac{\Gamma_2}{2}}{(i\Gamma + \omega_{2+})(i\Gamma + \omega_{2-})} e^{-\Gamma t} \bigg|^2
\end{split}
\end{equation}
and
\begin{equation}
\label{emx1}
\begin{split}
P_{\text{emx1}} (t, \Delta_1) &= g_{12}^{2}\frac{\Gamma_1}{2\pi} \bigg| \frac{e^{-i\omega_{1+}t}}{(\omega_{1+} - \omega_{1-})(\omega_{1+} + i \Gamma)}\\
&+ \frac{e^{-i\omega_{1-} t}}{(\omega_{1-} - \omega_{1+})(\omega_{1-} + i\Gamma)}\\
&+ \frac{e^{-\Gamma t}}{(i\Gamma + \omega_{1+})(i\Gamma + \omega_{1-})} \bigg|^2,\\
\end{split}
\end{equation}
where the eigenenergies of dressed states are $\omega_{1\pm} = \omega_1 - \omega_{0} - i\frac{\Gamma}{2} \pm \frac{1}{2} \sqrt{4 g_{12}^2 + (\Delta \omega_0 - i\frac{\Delta \Gamma}{2})^2}$ for $P_{\text{emx1}}$ and $\omega_{2\pm} = \omega_2 - \omega_{0} - i\frac{\Gamma}{2} \pm \frac{1}{2} \sqrt{4 g_{12}^2 + (\Delta \omega_0 - i\frac{\Delta \Gamma}{2})^2}$ for $P_{\text{em2}}$ respectively. Here $\omega_{0}=\frac{1}{2} (\omega_{01} + \omega_{02})$, $\Gamma = \frac{1}{2} (\Gamma_1 + \Gamma_2)$, $\Delta \omega_0 = \omega_{01} - \omega_{02}$ and $\Delta \Gamma= \Gamma_1 - \Gamma_2$.

Now we study the spontaneous emission of qubit 1 for both weak and strong dipole-dipole couplings. The parameters used are $g_{12} = 0.5 \Gamma_1$ for weak coupling and $g_{12} = 5 \Gamma_1$ for strong coupling. The time evolution of the survival probability of qubit 1, $P_{\text{total}} (t)$ is plotted in Figs. \ref{fig:purcell_rabi_time(c)} and \ref{fig:purcell_rabi_time(d)}. Here the blue dashed line is the SE in the free space for comparison. Both the Purcell enhancement and inhibition are observed in the system; the SE of qubit 1 is enhanced when $\Gamma_2 > \Gamma_1$  while inhibited when $\Gamma_2 < \Gamma_1$ (red solid curve). This is true for both weak and strong coupling. The inhibition of SE originates from the ``charger" effect which happens when the qubit 2 is in excited state and has lower decay rate than qubit 1. In the weak coupling regime, the enhancement is not significant, in contrast to the very large enhancement present in strong coupling regime. For both cases however, increasing $\Gamma_2$ to extremely large values will push qubit 2 into the bad-atom regime and cause the correlation between the qubits to decay exponentially; the SE of qubit 1 will approach the free space case. Additionally, by fine tuning the decay rates to $\Gamma_1 = \Gamma_2$, the two qubits are identical, and the survival probability of qubit 1 is not affected by the excited qubit 2, i.e. $P_{\text{total}} (t) = e^{-\Gamma_1 t}$ \cite{Liu:1975aa}. Therefore, in the strong coupling regime, the SE of qubit 1 can be significantly manipulated by the decay rate of qubit 2. Comparing with qubit 2 (the EME) prepared in the ground state in Fig. \ref{fig:purcell_rabi_time(b)}, the inhibition effect of qubit 1 is much stronger when qubit 2 is initially prepared in the excited state.

\begin{figure}
\subfigure{\includegraphics[width=4cm]{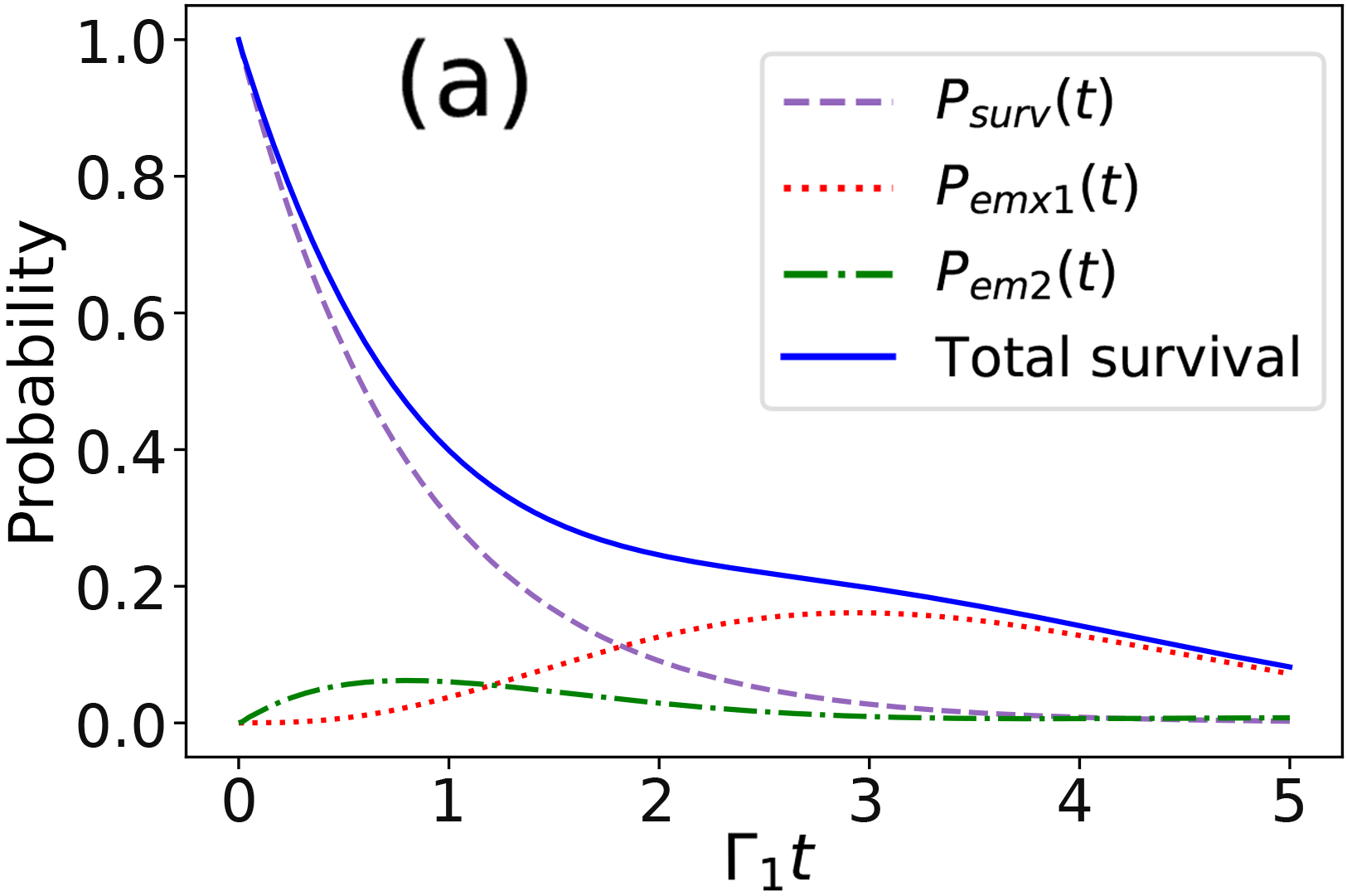}
\label{fig:|ee>_contribution(a)}}
\subfigure{\includegraphics[width=4cm]{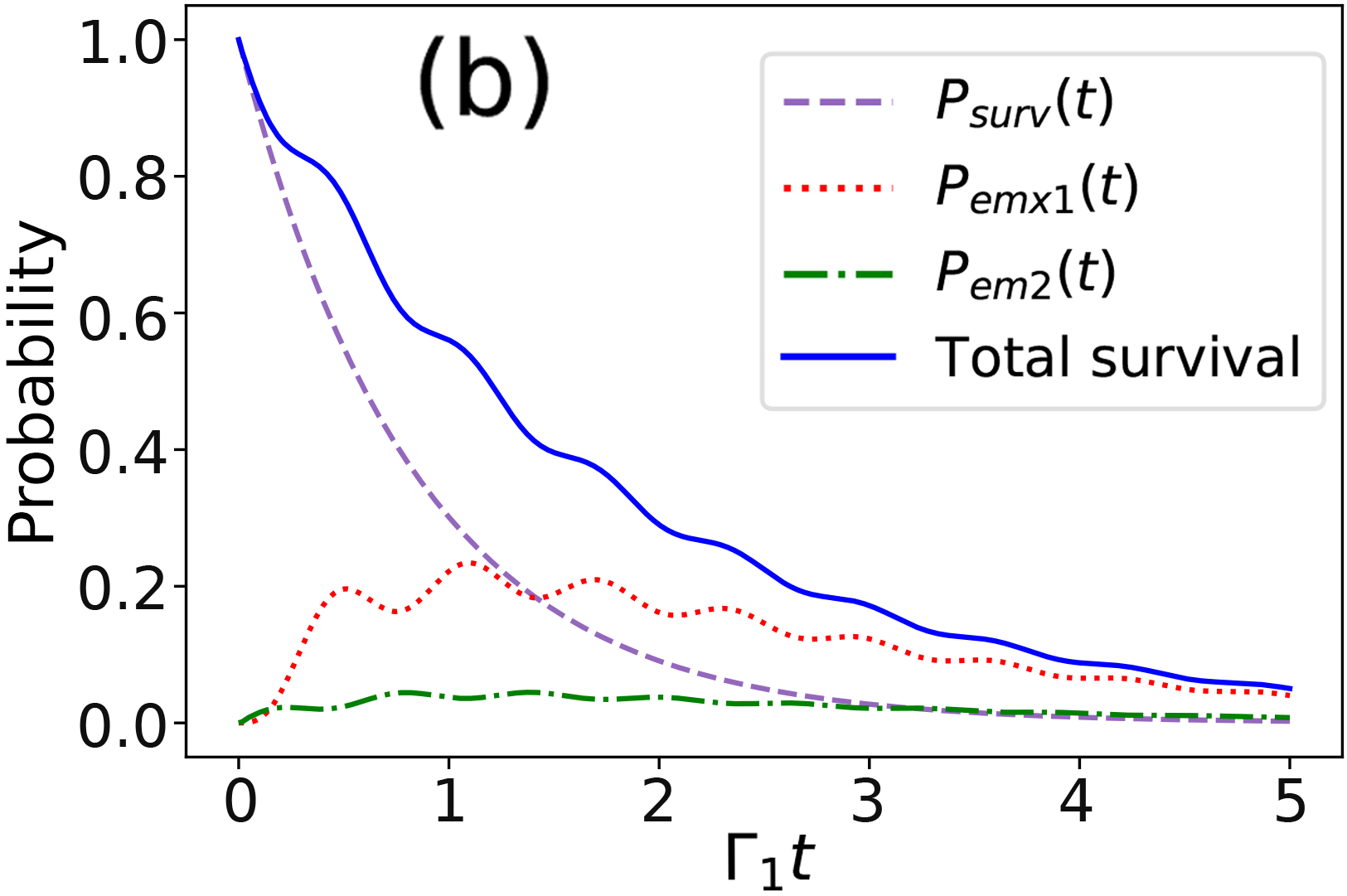}
\label{fig:|ee>_contribution(b)}} \\
\subfigure{\includegraphics[width=4cm]{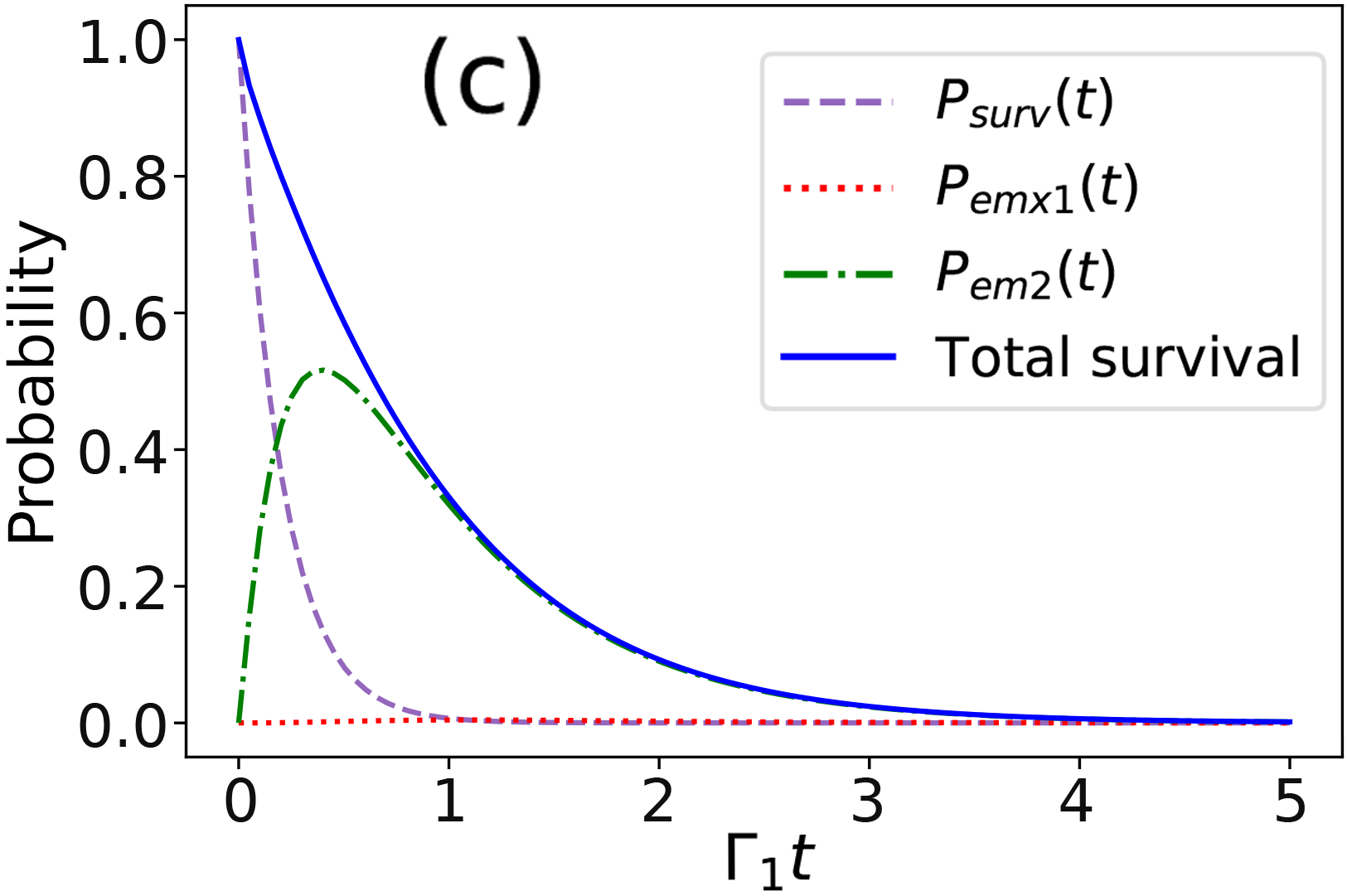}
\label{fig:|ee>_contribution(c)}}
\subfigure{\includegraphics[width=4cm]{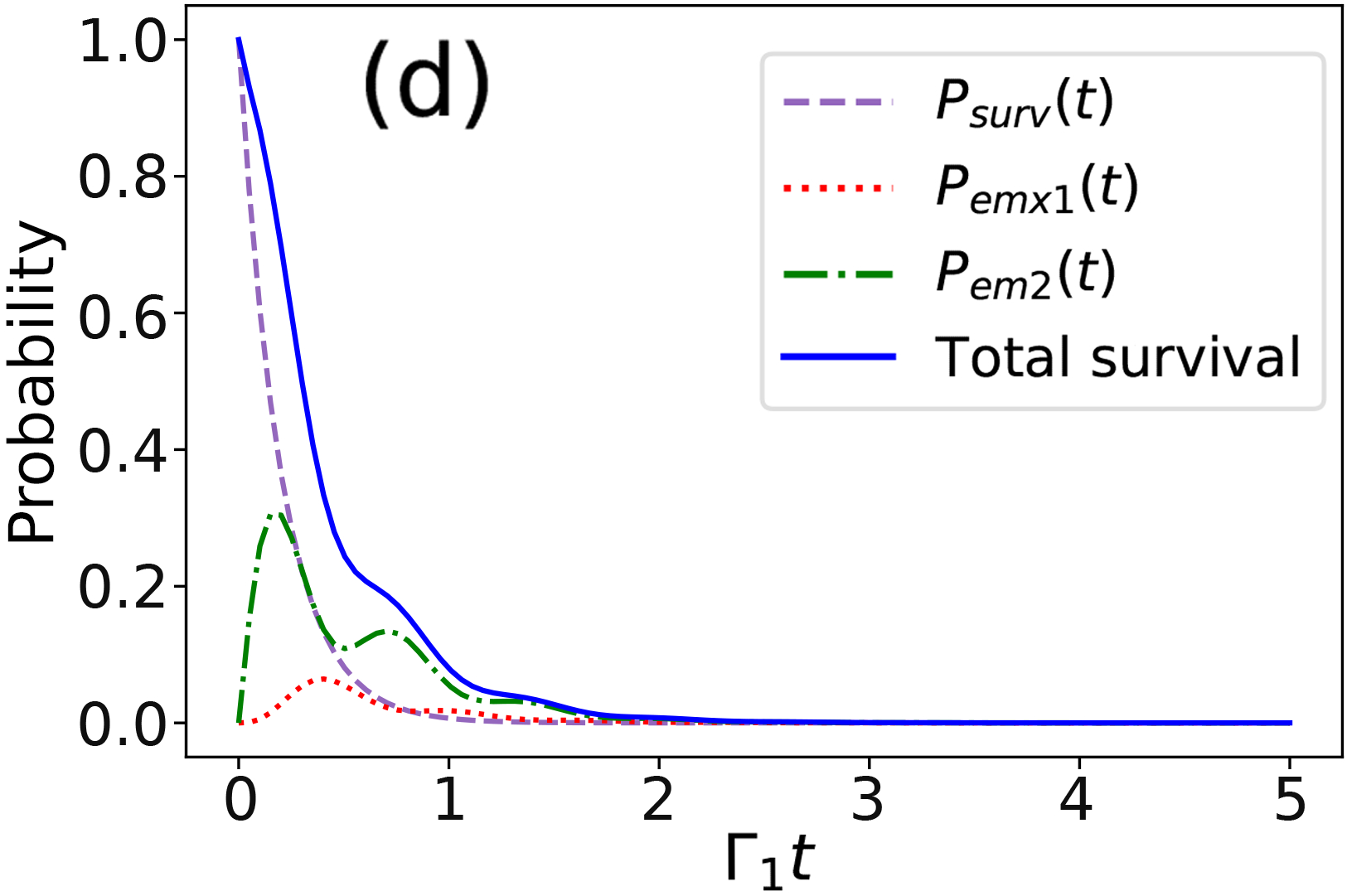}
\label{fig:|ee>_contribution(d)}}
\caption{(color online). Various contributions for the survival probability of state $|e_{1}\rangle$ (blue solid) in weak and strong coupling regimes. The purple dashed, green dash-dotted and red dotted curves are for the survival probability of $|e_{1}e_{2}\rangle$, the single-photon emission probability via qubit 2, and the emission-exchange probability in Eq. (\ref{Pem2_2qubits}-\ref{emx1}). The parameters are (a) $g_{12} = 0.5\Gamma_1, \Gamma_2 = 0.2\Gamma_1$; (b) $g_{12} = 5\Gamma_1, \Gamma_2 = 0.2\Gamma_1$; (c) $g_{12} = 0.5\Gamma_1, \Gamma_2 = 4\Gamma_1$; (d) $g_{12} = 5\Gamma_1, \Gamma_2 = 4\Gamma_1$.}
\label{fig:|ee>_contribution}
\end{figure}

To further understand the oscillating behaviour in the Purcell inhibition regime in Fig. \ref{fig:purcell_rabi_time(d)}, we study the three contributions to $P_{\text{total}} (t)$. For $\Gamma_2 = 0.2\Gamma_1$ in Fig. \ref{fig:|ee>_contribution(a)} and \ref{fig:|ee>_contribution(b)}, the spontaneous emission of qubit 1 is suppressed. In the strong coupling regime ($g_{12} = 5\Gamma_1$), the dipole-dipole coupling sets up a Rabi oscillation in the system, hence causing $P_{\text{total}} (t)$ to exhibit oscillatory behaviour. The oscillating effect is mainly due to $P_{\text{emx1}} (t)$ which directly involves the coupling process $g_{12}$ as shown in Eq. (\ref{emx1}). Due to the small $\Gamma_2$, the contribution from $P_{\text{em2}} (t)$ is not as significant. The oscillation is small because both qubits are initially in the excited state, making it difficult for Rabi oscillation to form at initial times. This is contrasted with the large vacuum Rabi oscillation in the single-excitation case plotted in Fig. \ref{fig:purcell_rabi_time(a)} and \ref{fig:purcell_rabi_time(b)}, where qubit 2 is initially in the ground state.
In comparison, for large $\Gamma_2 = 4\Gamma_1$ in Fig. \ref{fig:|ee>_contribution(c)} and \ref{fig:|ee>_contribution(d)}, Purcell enhancement is observed. Due to large $\Gamma_2$, the contribution from $P_{\text{emx1}} (t)$ is not as significant. The enhancement effect is mainly due to $P_{\text{em2}} (t)$. In the weak coupling regime, the emission-exchange probability $P_{\text{emx1}} (t)$ is negligible, hence there is little Purcell enhancement from the free-space case. In the strong coupling regime however, a significant Rabi oscillation between the qubits results in a large enhancement in spontaneous emission.

\begin{figure*}
\subfigure{\includegraphics[width=5.3cm]{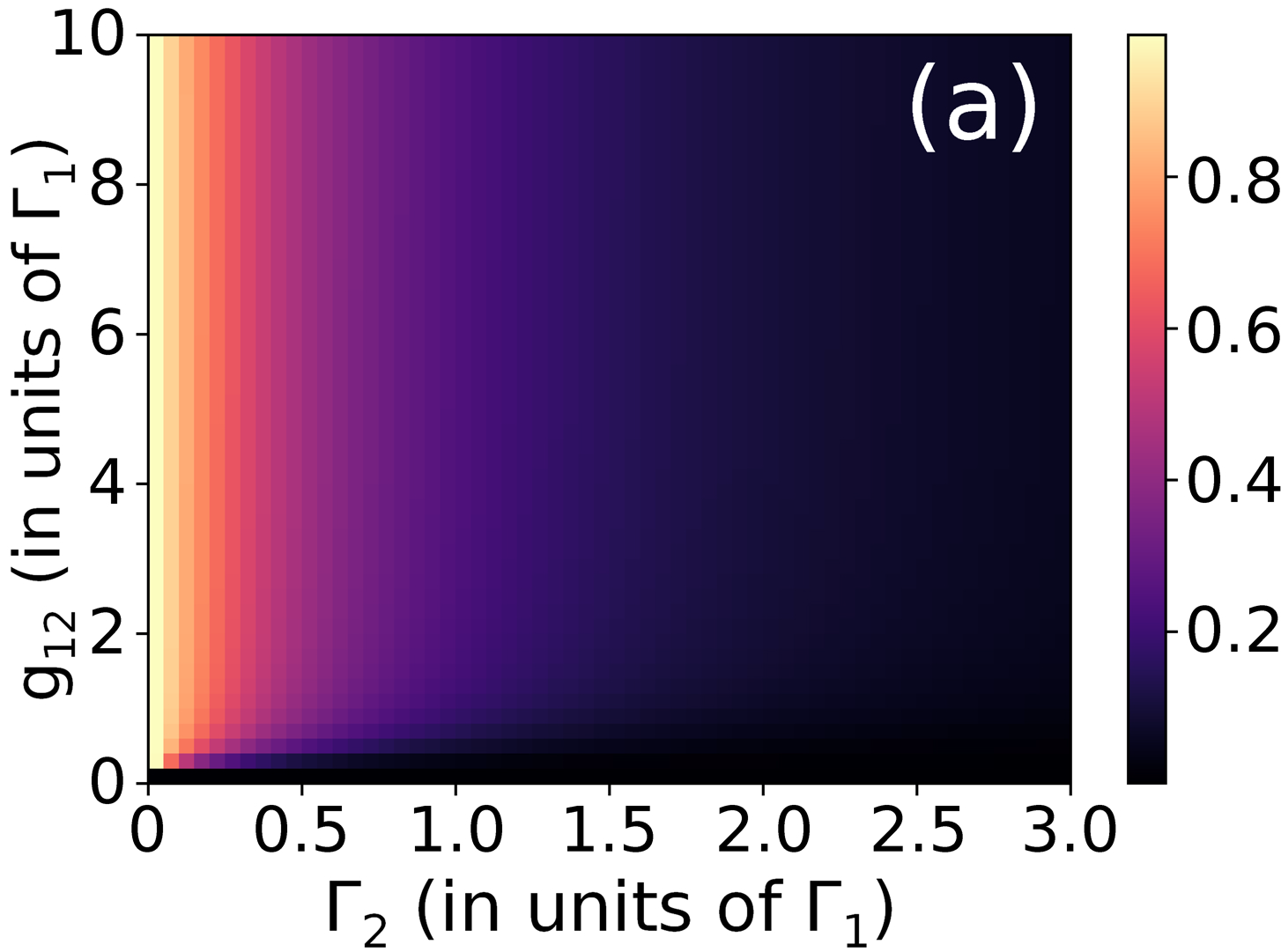}
\label{fig:twoem_|ee>_contour_a}}
\subfigure{\includegraphics[width=5.3cm]{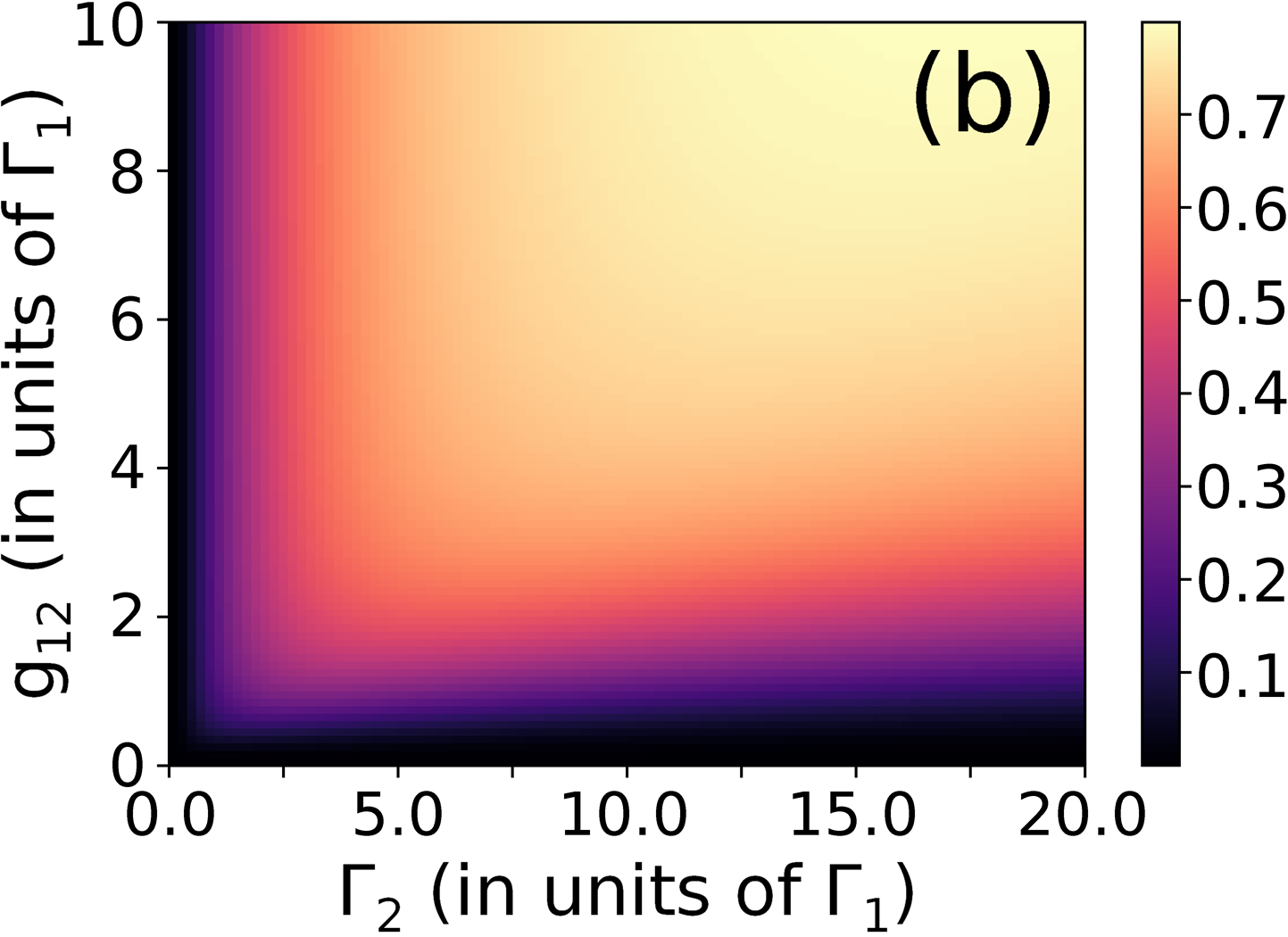}
\label{fig:twoem_|ee>_contour_b}}
\subfigure{\includegraphics[width=5.3cm]{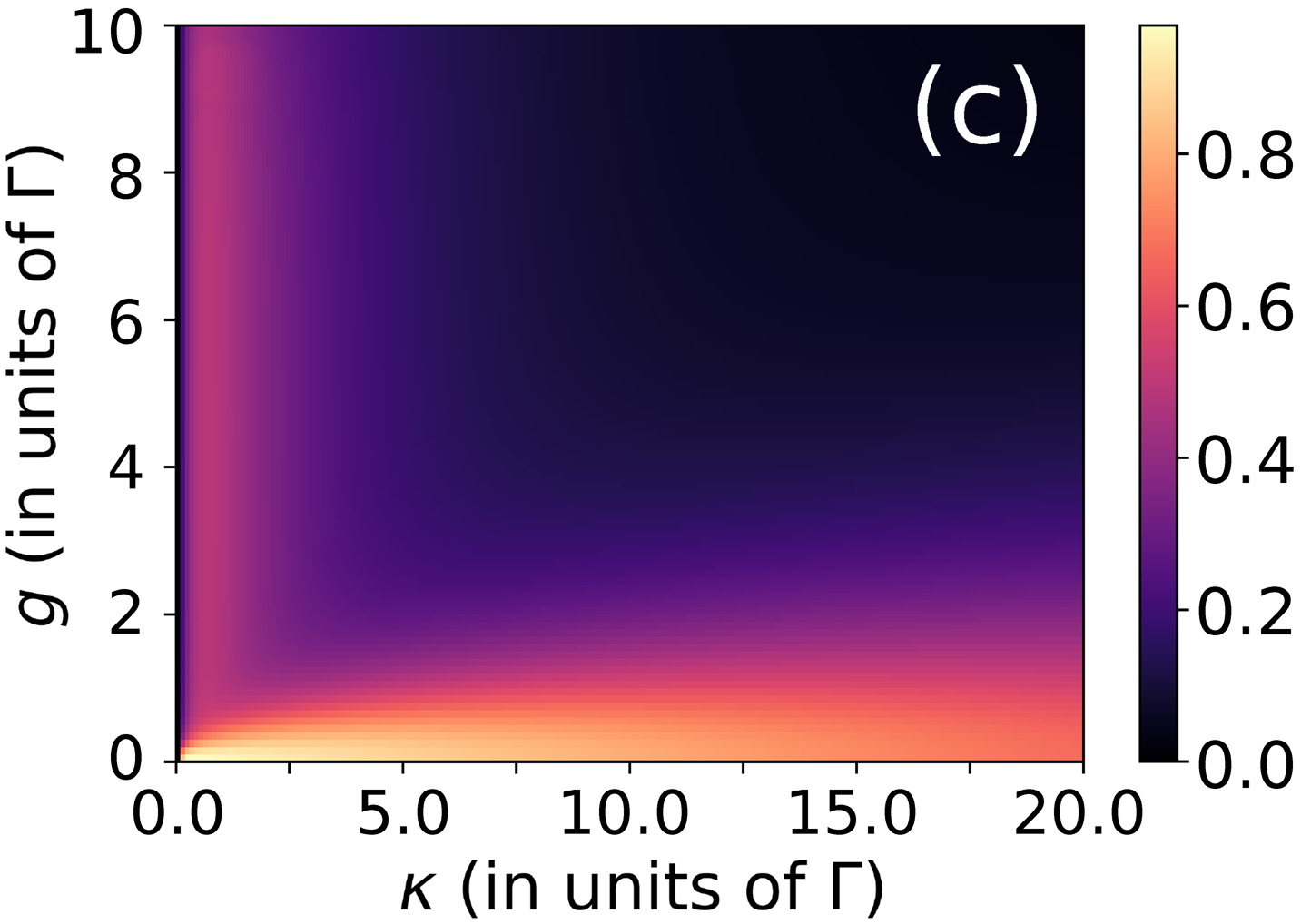}
\label{fig:twoem_|ee>_contour_c}}
\subfigure{\includegraphics[width=5.3cm]{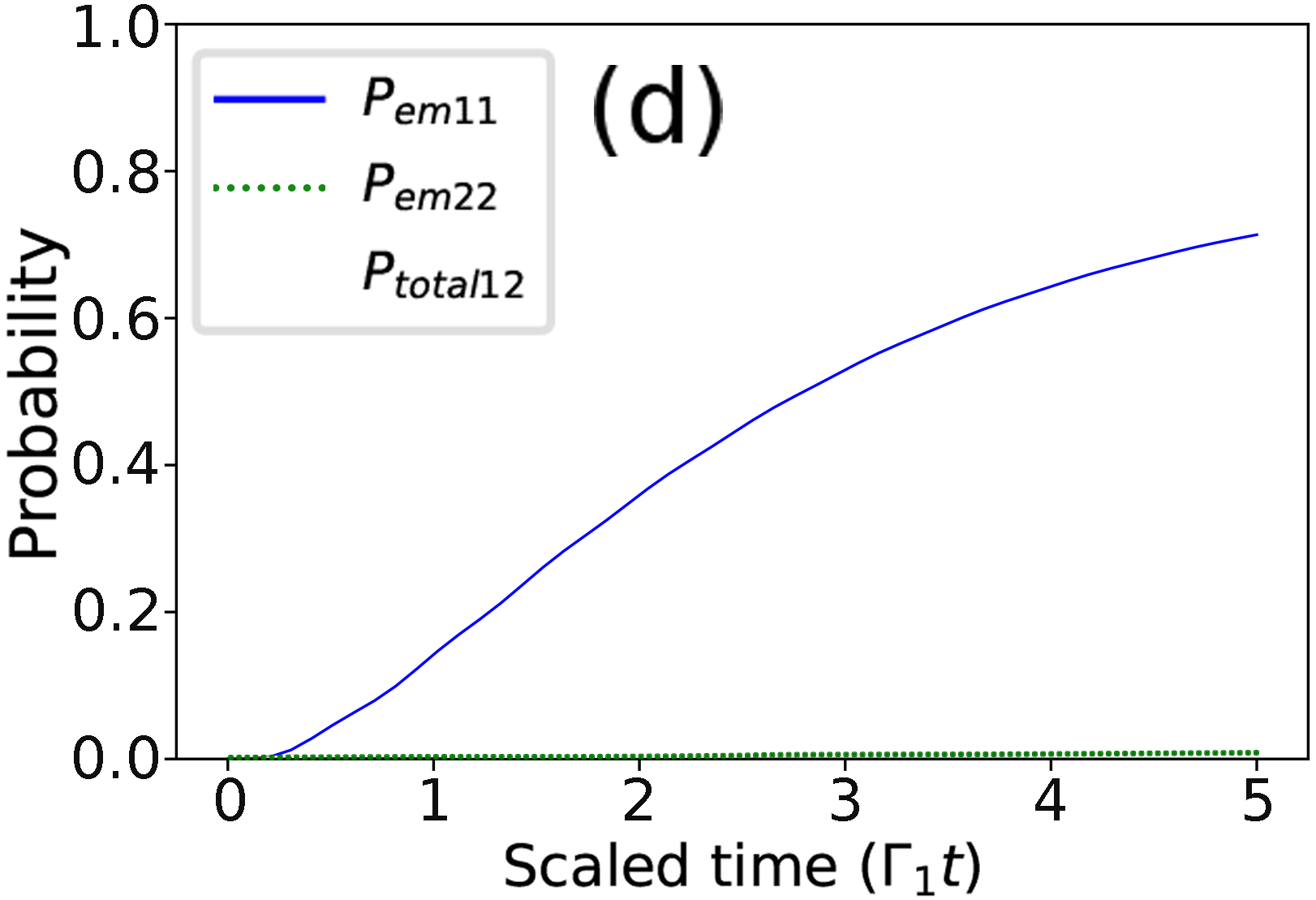}
\label{fig:twoem_|ee>_contour_d}}
\subfigure{\includegraphics[width=5.3cm]{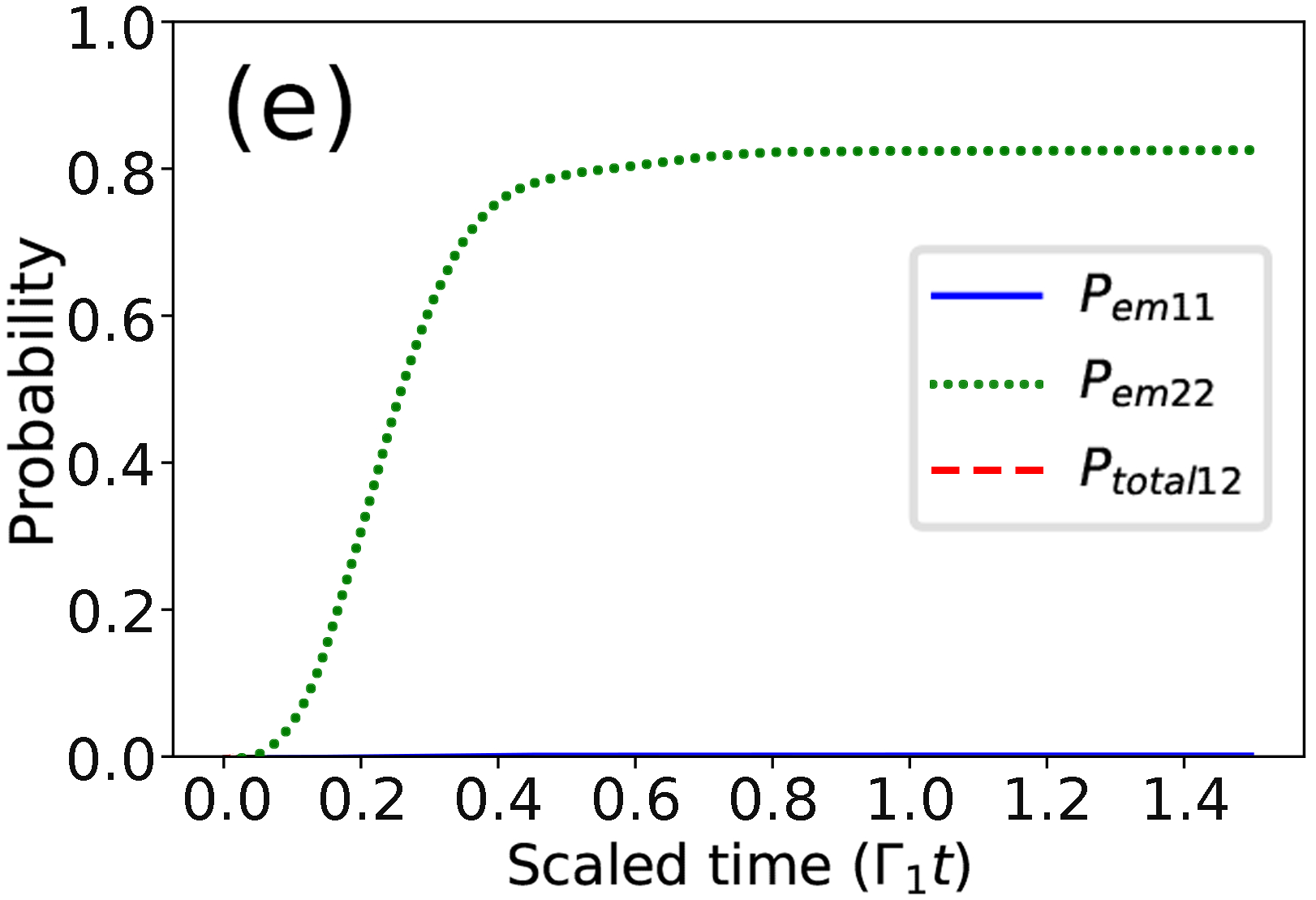}
\label{fig:twoem_|ee>_contour_e}}
\subfigure{\includegraphics[width=5.3cm]{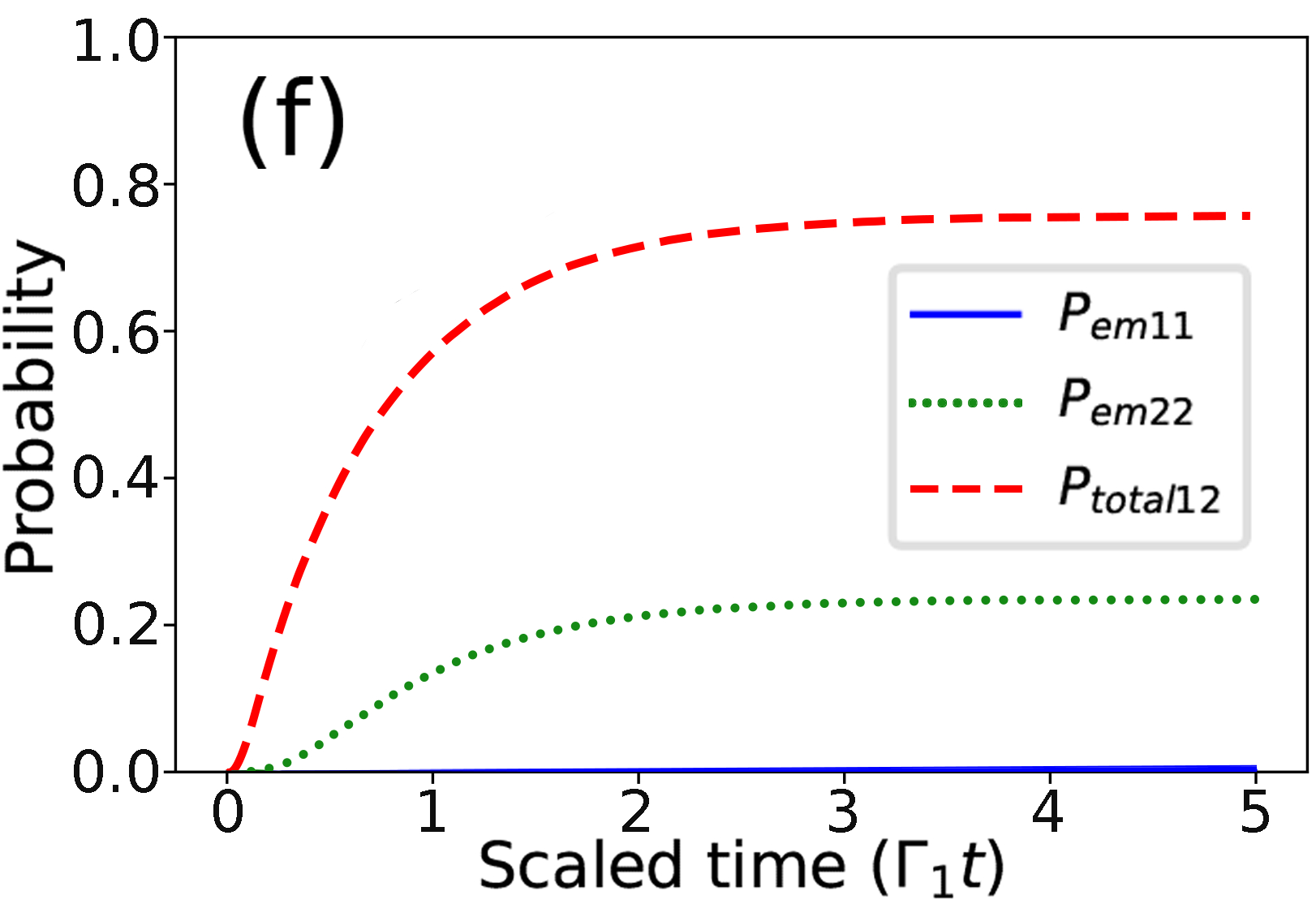}
\label{fig:twoem_|ee>_contour_f}}
\caption{(color online). Two-photon emission probabilities shown in Eq. (\ref{em11}-\ref{em21}) in steady state. (a) $P_{\text{em11}}$, the steady state is $b_1^\dag b_1^\dag |g_{1}g_{2}\rangle$; (b) $P_{\text{em22}}$, the steady state is $b_2^\dag b_2^\dag |g_{1}g_{2}\rangle$; (c) $P_{\text{total12}}$, the steady state is $b_1^\dag b_2^\dag |g_{1}g_{2}\rangle$. Atomic detuning $\Delta \omega_0$ is taken to be zero. Time evolution of two-photon emission probabilities. (d) $g_{12} = 6\Gamma_1, \Gamma_2 = 0.1\Gamma_1$. $P_{\text{em11}}$ is dominant, (e) $g_{12} = 8\Gamma_1, \Gamma_2 = 15\Gamma_1$. $P_{\text{em22}}$ is dominant, (f) $g_{12} = \Gamma_1, \Gamma_2 = 10\Gamma_1$. $P_{\text{total12}}$ is dominant.}
\label{fig:twoem_|ee>_contour}
\end{figure*}

We now show the controllability of decay channels for the SE of the system. When the subsystem of coupled qubits is exposed to the environments, it will eventually decay to its ground state $|{g_{1}g_{2}}\rangle$ through various emission routes. During this process, there are three sub-processes involved: $P_{\text{em11}}(t) = \iint_{-\infty}^\infty \mathrm{d}\Delta_1  \mathrm{d}\Delta_1^\prime \, P_{\text{em11}} (t, \Delta_1, \Delta_1^\prime)$, where both photons are emitted via qubit 1; $P_{\text{em22}} (t) = \iint_{-\infty}^\infty \mathrm{d}\Delta_2  \mathrm{d}\Delta_2^\prime \, P_{\text{em22}} (t, \Delta_2, \Delta_2^\prime)$, where both photons are emitted via qubit 2 and $P_{\text{total12}} (t)= \iint_{-\infty}^\infty \mathrm{d} \Delta_1 \mathrm{d} \Delta_2 [P_{\text{em12}} (t, \Delta_1, \Delta_2) + P_{\text{em21}} (t, \Delta_1, \Delta_2)]$, where each qubit emits one photon. (For explicit expressions of the frequency-resolved propagators, please refer to Appendix \ref{double_excitation}). By studying the probabilities of these two-photon emission sub-processes in different parameters, it may be useful in optimising the desired emission route or to suppress unwanted emissions for applications. Here we first study these probabilities in the long-time limit. From Fig. \ref{fig:twoem_|ee>_contour_a} to \ref{fig:twoem_|ee>_contour_c}, it can be seen that in spite of the dipole-dipole interaction strength $g_{12}$, the probability $P_{\text{em11}}$ is greatest due to small decay rate $\Gamma_2$ of qubit 2. In contrast, when both $g_{12}$ and $\Gamma_2$ is large, $P_{\text{em22}}$ is large. The final state has a high emission probability through qubit 2 and decays in state $b_2^\dag b_2^\dag |{g_{1}g_{2}}\rangle$. Under weak dipole-dipole coupling, $P_{\text{total12}}$ dominates the emission process. The corresponding time-evolutions of the two-photon emission probabilities are also shown from Fig. \ref{fig:twoem_|ee>_contour_d} to \ref{fig:twoem_|ee>_contour_f}. For example, in Fig. \ref{fig:twoem_|ee>_contour_d}, $P_\text{em11}$ is the dominant probability when $g_{12} = 6\Gamma_1, \Gamma_2 = 0.1\Gamma_1$. Similarly, $P_\text{em11}$ dominates when $g_{12} = 8\Gamma_1, \Gamma_2 = 15\Gamma_1$ as in Fig. \ref{fig:twoem_|ee>_contour_e}, and $P_\text{total12}$ dominates when $g_{12} = \Gamma_1, \Gamma_2 = 10\Gamma_1$ as in Fig. \ref{fig:twoem_|ee>_contour_f}. The time-evolution results are consistent with the steady-state results from Fig. \ref{fig:twoem_|ee>_contour_a} to \ref{fig:twoem_|ee>_contour_c}. We summarise the dominant emission process in various regimes in Tab. (\ref{sumtab}) which might find its use in controlling the emission routes for applications.

\begin{table}[htp]
\centering
\begin{tabular}{|c|l|l|}
\hline
\multicolumn{1}{|l|}{}  & \multicolumn{1}{c|}{\textbf{Small $\Gamma_2$}} & \multicolumn{1}{c|}{\textbf{Large $\Gamma_2$}} \\ \hline
\textbf{Small $g_{12}$} & $b_1^\dag b_2^\dag |{g_1 g_2}\rangle$                   & $b_1^\dag b_2^\dag |{g_1 g_2}\rangle$                   \\ \hline
\textbf{Large $g_{12}$} & $b_1^\dag b_1^\dag |{g_1 g_2}\rangle$                   & $b_2^\dag b_2^\dag |{g_1 g_2}\rangle$                   \\ \hline
\end{tabular}
\caption{Quantum states with the largest probability in different regimes.}
\label{sumtab}
\end{table}

\section{Single qubit coupled to a cavity: two excitations}
\label{sec:JCM_n=1}

Here we put the qubit 1 into a cavity and study the influence to the SE of the qubit 1. Thus we set $g_{2}=g_{12}=\omega_{02}=\Gamma_{2}=0$ in Eq. (\ref{H0}) to Eq. (\ref{H01}); the whole system simplifies to the well-known Jaynes-Cummings model. In our study, the cavity is initially prepared either in the ground state $|0\rangle$ or Fock state $|{1}\rangle$.
The effect of the cavity in ground state $|0\rangle$ is equivalent to the two-qubit system with qubit 2 in the ground state $|{g_{2}}\rangle$. Thus, the phenomena discussed in Sec. \ref{two_qubits_single_excitation} (Purcell effect, Rabi oscillations, control of decay channel) are present here as well.

\begin{figure}
\includegraphics[width=9cm]{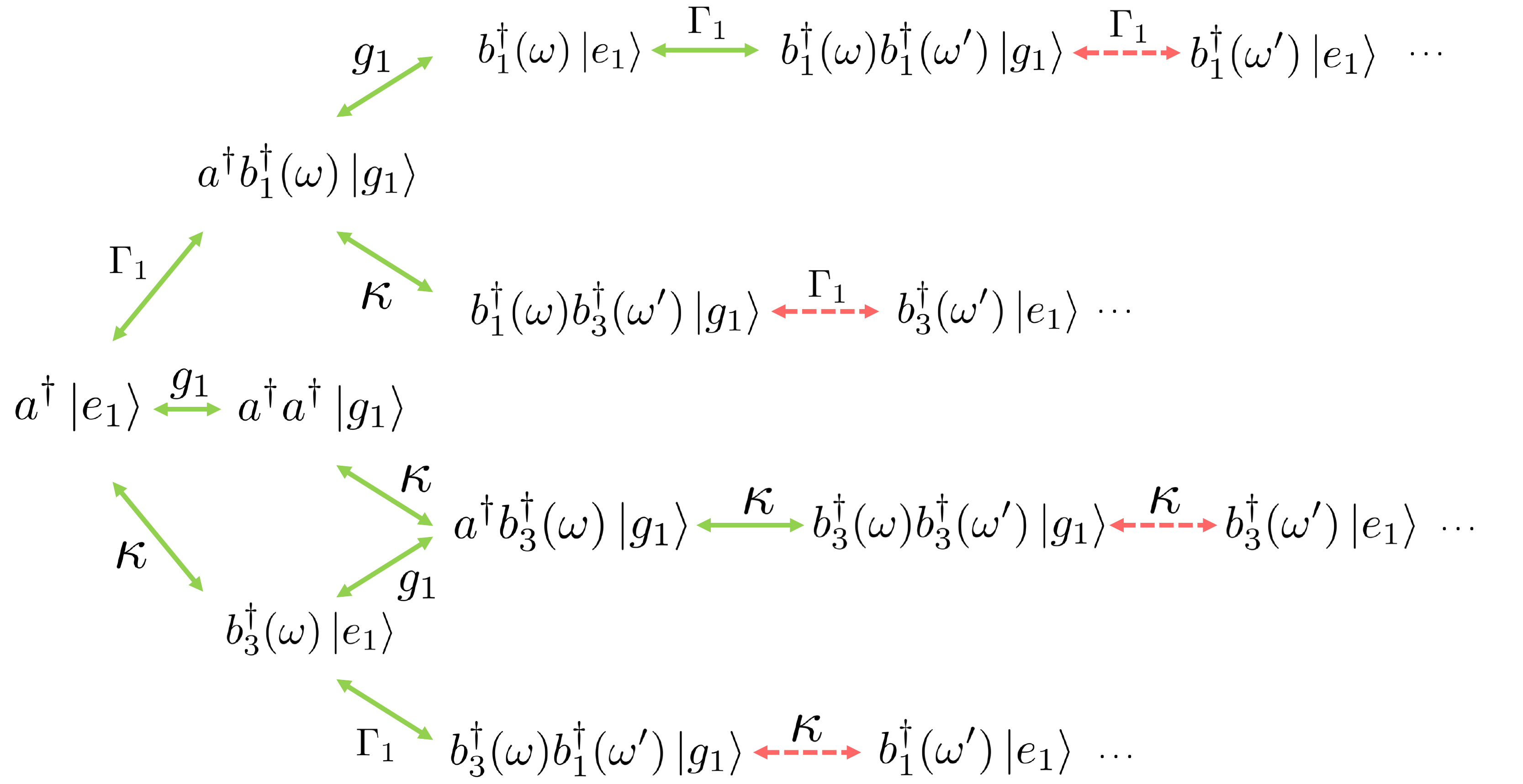}
\caption{(color online). Network of quantum states for single excited qubit in a cavity with Fock state $|1\rangle$. The network is terminated when the system reaches ground state, indicated by the dashed arrows.}
\label{fig:JC2_network}
\end{figure}

For the double excitation case, we initially set the cavity in the Fock state $|{1}\rangle$ so that the initial state of the system is $a^\dag |{e_{1}}\rangle$. By the similar method as shown in Eq. (\ref{EOM}), we obtain a closed differential system of propagators for calculations with the assumption that the network of quantum states shown in Fig. \ref{fig:JC2_network} is terminated whenever both the qubit and cavity reach ground state. The cut-off of the network is depicted by the red dashed arrows in Fig. \ref{fig:JC2_network}.

From the network of quantum states, we find that compared with the two-qubit double excitation system, there is a new branch of states with double occupation of cavity mode, i.e., Fock state $|n=2\rangle$ during the time evolution of the system. This makes the SE of qubit 1 different from the two-qubit system with two excitations. Fig. \ref{fig:purcell_rabi_time(e)} and \ref{fig:purcell_rabi_time(f)} show the transition from Purcell inhibition to enhancement as $\kappa$ is increased. Comparing with the single-excitation JCM in Fig. \ref{fig:purcell_rabi_time(a)} and \ref{fig:purcell_rabi_time(b)}, the inhibition effect is enhanced since the survival probability only dips below the free-space value once, and stays above the blue solid curve for the rest of the time. Comparing with the two-qubit double excitations system in Fig. \ref{fig:purcell_rabi_time(c)} and \ref{fig:purcell_rabi_time(d)}, the inhibition effect of double-excitation JCM is suppressed. The previous work by Sete et al. \cite{PhysRevB.89.104516} considered the suppression of spontaneous emission in JCM by detuning the resonator with the qubit, thus decreasing their spectral overlap. In this work, we further conclude the inhibition effect due to differences in the decay rates between the qubit and cavity in both strong and weak coupling regimes.

\section{Two qubits coupled to a cavity: single excitation}
\label{sect:eg+cavity}

In previous sections, we have found that a single excited qubit in a cavity (JCM) exhibits both Purcell enhancement and inhibition, and that the two-qubit system contains identical behaviour as JCM with initial Fock state $|n=0\rangle$. In the general system of Eq. (\ref{H0}), there are three couplings involved: $g_1$ between qubit 1 and cavity, $g_2$ between qubit 2 and cavity, and $g_{12}$ between the qubits. Whether these couplings are cooperative or competitive to the spontaneous emission of qubit 1 is our topic of study here. Note that the present system can be regarded as a Tavis-Cummings model \cite{Chen:2009aa}. In this section, we study the system in both time domain in Sec. \ref{sec:TCM_time} and frequency domain in Sec. \ref{sect:spectral}.

\subsection{Time evolution of transition probabilities}
\label{sec:TCM_time}
\begin{figure}
\includegraphics[width=9cm]{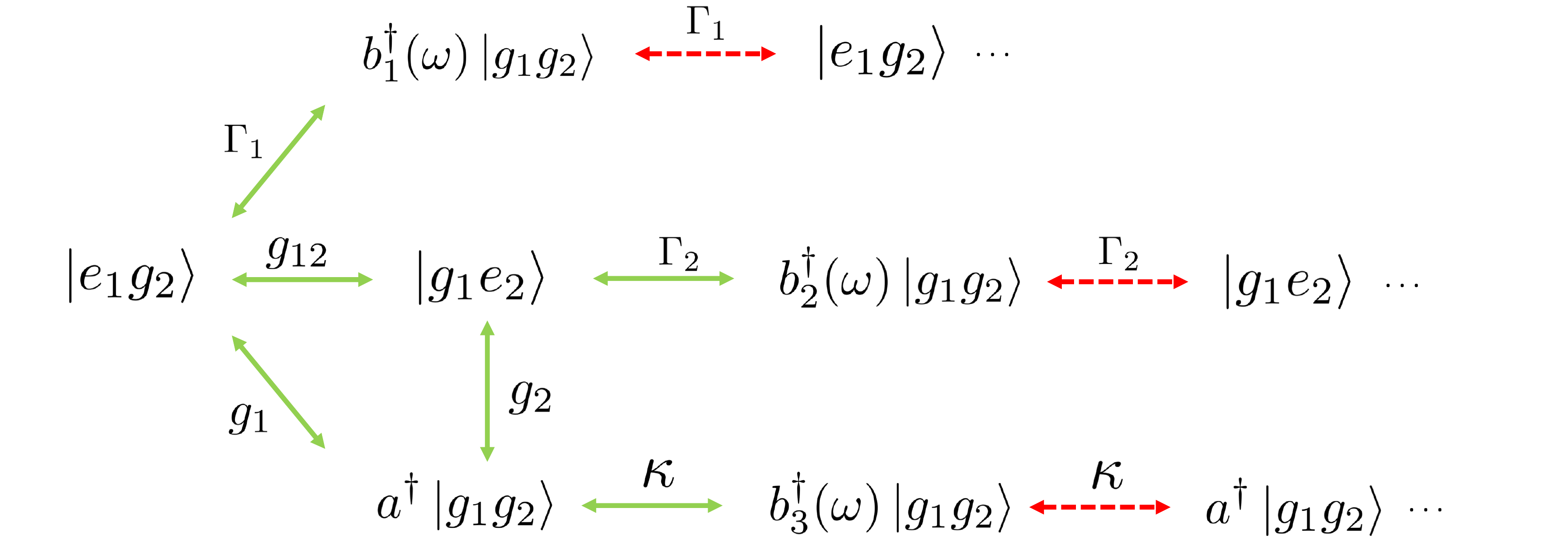}
\caption{(color online). Network of quantum states for two qubits in a cavity with Fock state $|0\rangle$. The two qubits are coupled via dipole-dipole interactions. The network is terminated when the system reaches ground state, indicated by the dashed arrows.}
\label{fig:eg+c_network}
\end{figure}

By the similar method as shown in Eq. (\ref{EOM}), we obtain a closed differential system of propagators for calculations. The network of quantum states is shown in Fig. \ref{fig:eg+c_network}. The expressions for survival probability of qubit 1, $P_\text{surv}(t) = | \langle{e_1 g_2}| U(t) | e_1 g_2 \rangle|^2$; the spontaneous emission probability of qubit 1, $P_\text{em1}(t, \Delta_1) = | \langle{g_1 g_2}| b_1 (\omega_1) U (t) | e_1 g_2 \rangle |^2$; the exchange-emission probability via qubit 2, $P_\text{emx2}(t, \Delta_2) = | \langle{g_1 g_2}| b_2 (\omega_2) U(t) | e_1 g_2 \rangle |^2$ and the exchange-emission probability via the cavity, $P_\text{emr}(t, \Delta_r) = | \langle{g_1 g_2} b_3 (\omega_r) U(t) | e_1 g_2 \rangle |^2$ are given in Appendix \ref{appendix_eg+c}.

It is observed that the spontaneous emission of qubit 1 can be either enhanced or inhibited as shown in Fig. \ref{fig:purcell_rabi_time(g)} and Fig. \ref{fig:purcell_rabi_time(h)}, respectively. As discussed in previous sections, the condition for inhibition in JCM is $\kappa < \Gamma_1$, or $\Gamma_2 < \Gamma_1$ in the equivalent two-qubit system. In the current system however, slight inhibition is still observed even if $\Gamma_2 > \Gamma_1$, provided $\kappa$ is sufficiently smaller than $\Gamma_1$.

\begin{figure}
\subfigure{\includegraphics[width=4cm]{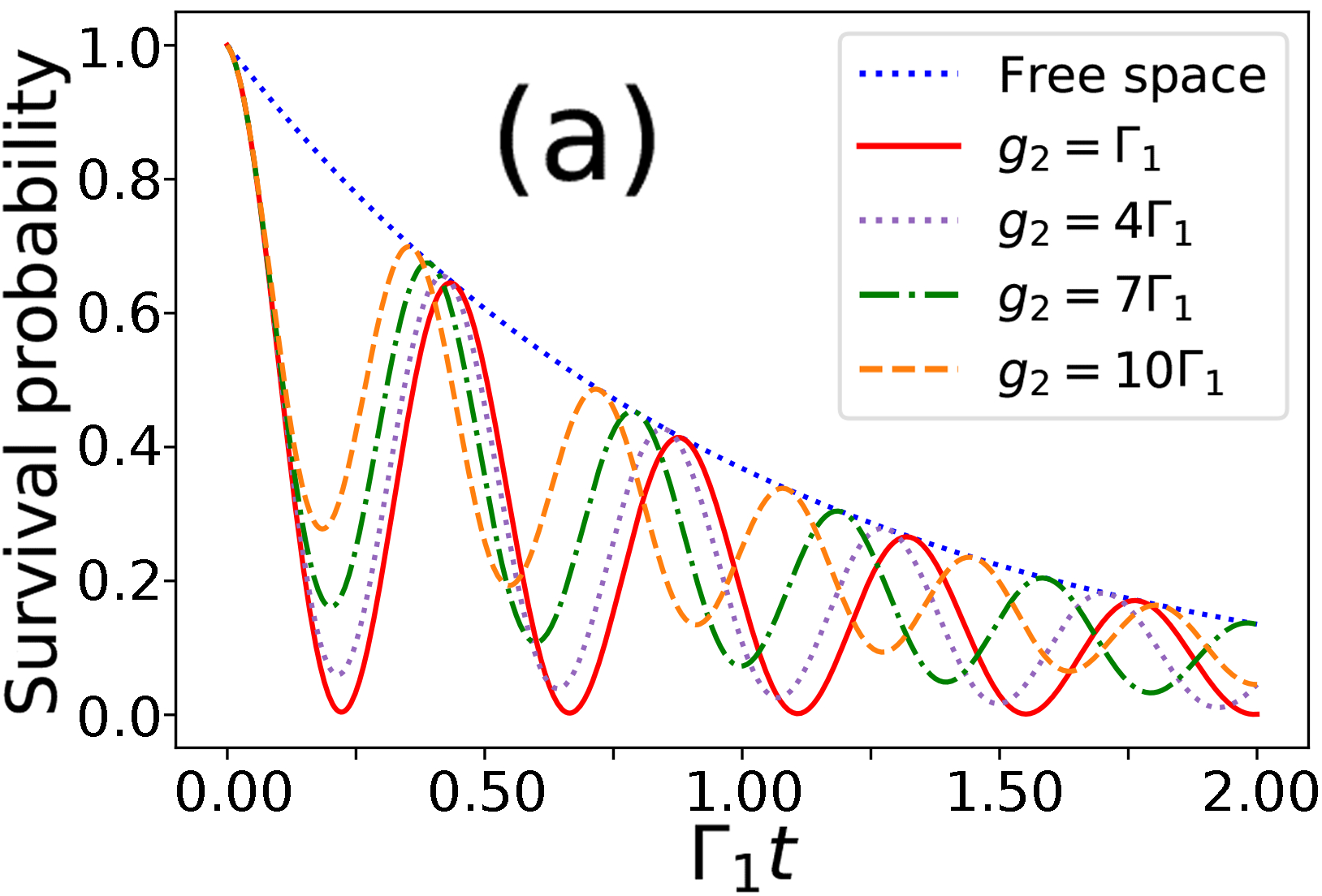}
\label{fig:|eg>+c_g2detuning(a)}}
\subfigure{\includegraphics[width=4cm]{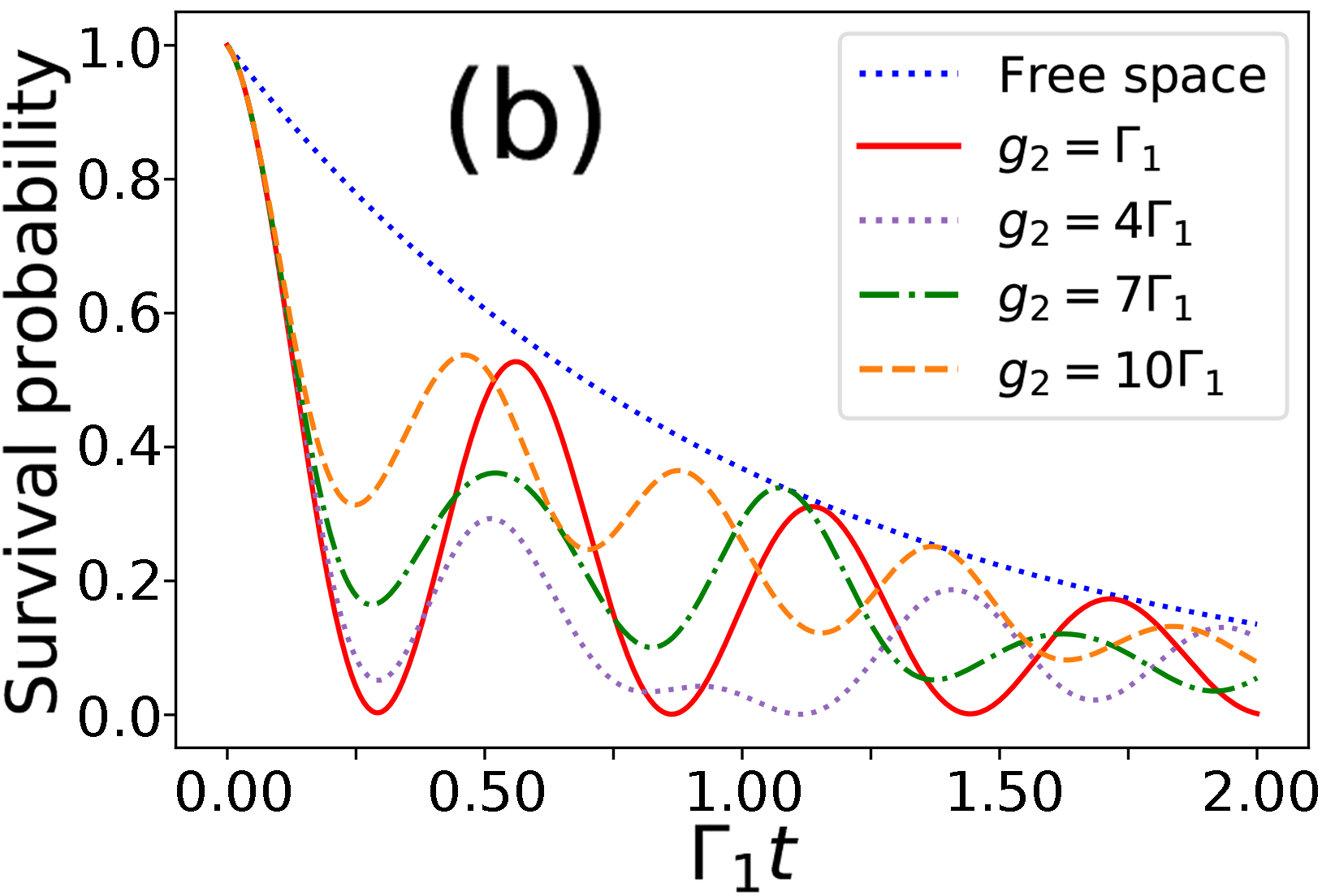}
\label{fig:|eg>+c_g2detuning(b)}}
\caption{(color online). Survival probability in Eq. (\ref{A2}) of qubit 1 for illustrating the detuning effect of $g_{2}$. The parameters are $\kappa = \Gamma_2 = \Gamma_1$ and (a) $g_1 = 5\Gamma_1, g_{12} = 5\Gamma_1$ ($g_1 = g_{12}$); (b) $g_1 = 5\Gamma_1, g_{12} = 2\Gamma_1$ ($g_1 \neq g_{12}$).}
\label{fig:|eg>+c_g2detuning}
\end{figure}

In Fig. \ref{fig:purcell_rabi_time(g)}, it can be seen that when $g_2$ is the dominant coupling (green dashed-dotted curve), the survival probability approaches the free-space case (blue solid curve). This can be explained as follows: For $g_2 \gg (g_1, g_{12})$, when qubit 1 emits into the cavity, the cavity photon is most likely to be absorbed by qubit 2 to set up a Rabi oscillation between qubit 2 and the cavity. Hence, the Rabi oscillation between qubit 1 and cavity is suppressed. Moreover, the cavity photon does not get absorbed back by qubit 1 easily since $g_2$ is dominant. As a result, qubit 1 effectively decouples from the system, and the emission behaviour favours spontaneous emission via $\Gamma_1$. The decoupling effect of $g_2$ is seen more clearly in Fig. \ref{fig:|eg>+c_g2detuning}.
When $g_1 = g_{12}$, the oscillations are coherent. Increasing $g_2$ leads to greater effective decoupling between qubit 1 and its EME, and the survival probability of qubit 1 approaches the free-space curve. On the other hand, when $g_1 \neq g_{12}$, the oscillations become out of phase. Physically, $g_1$ and $g_{12}$ are both responsible for separate Rabi oscillations involving qubit 1. If $g_1 \neq g_{12}$, the two Rabi frequencies are different, leading to a more complicated interference result. Also, in this case, the decoupling effect of $g_2$ is only significant if $g_2$ is the largest coupling constant, as shown in Fig. \ref{fig:|eg>+c_g2detuning(b)}. From our result, we can conclude that $g_1$ and $g_{12}$ are cooperative couplings significant to the enhancement or inhibition of spontaneous emission. On the other hand, $g_2$, which does not involve qubit 1 directly, serves as an effective decoupling parameter competing with $g_1$ and $g_{12}$.

\begin{figure*}
\centering
\subfigure{\includegraphics[width=5.3cm]{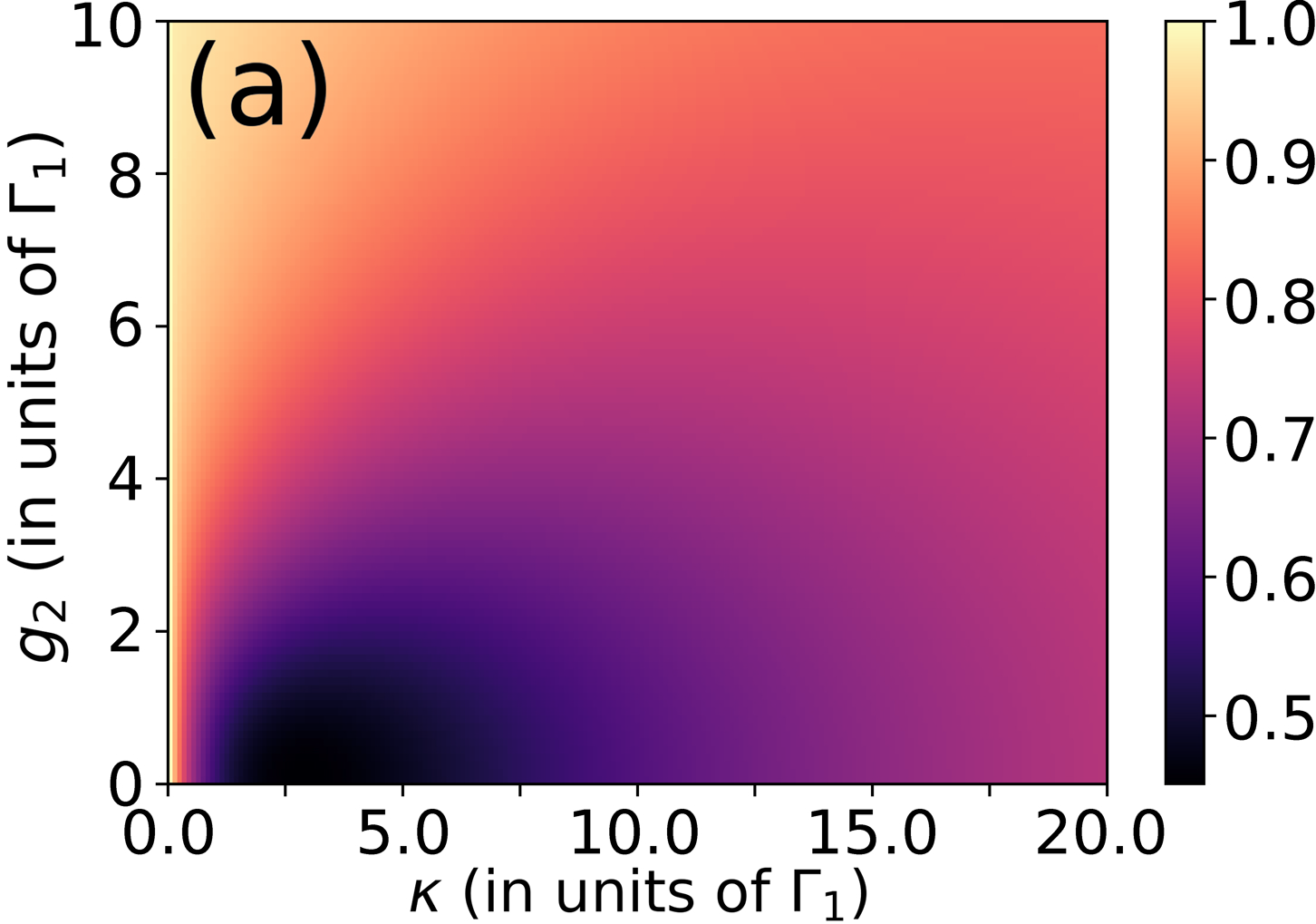}
\label{fig:|eg>+c_decayroute(a)}}
\subfigure{\includegraphics[width=5.3cm]{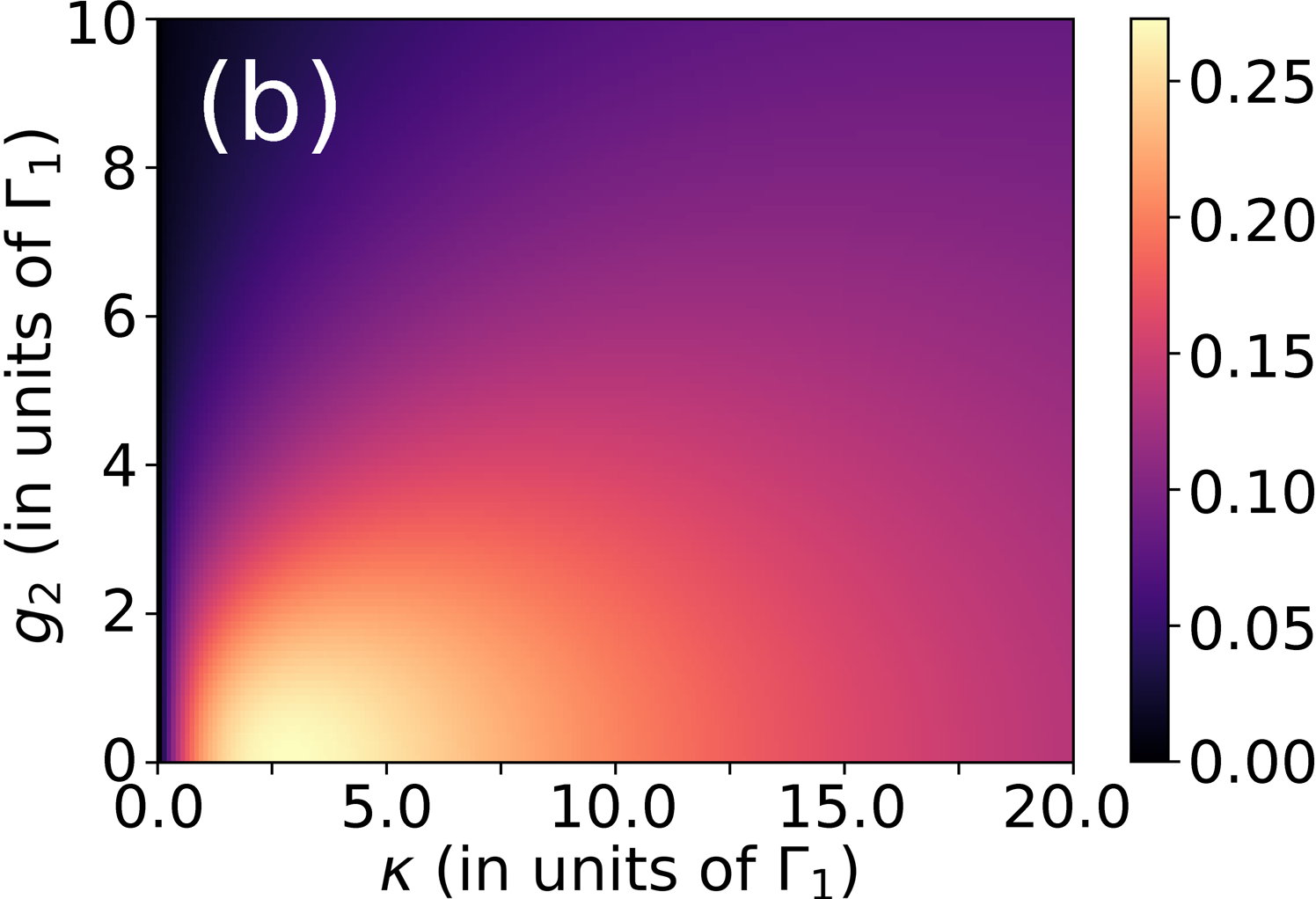}
\label{fig:|eg>+c_decayroute(b)}}
\subfigure{\includegraphics[width=5.3cm]{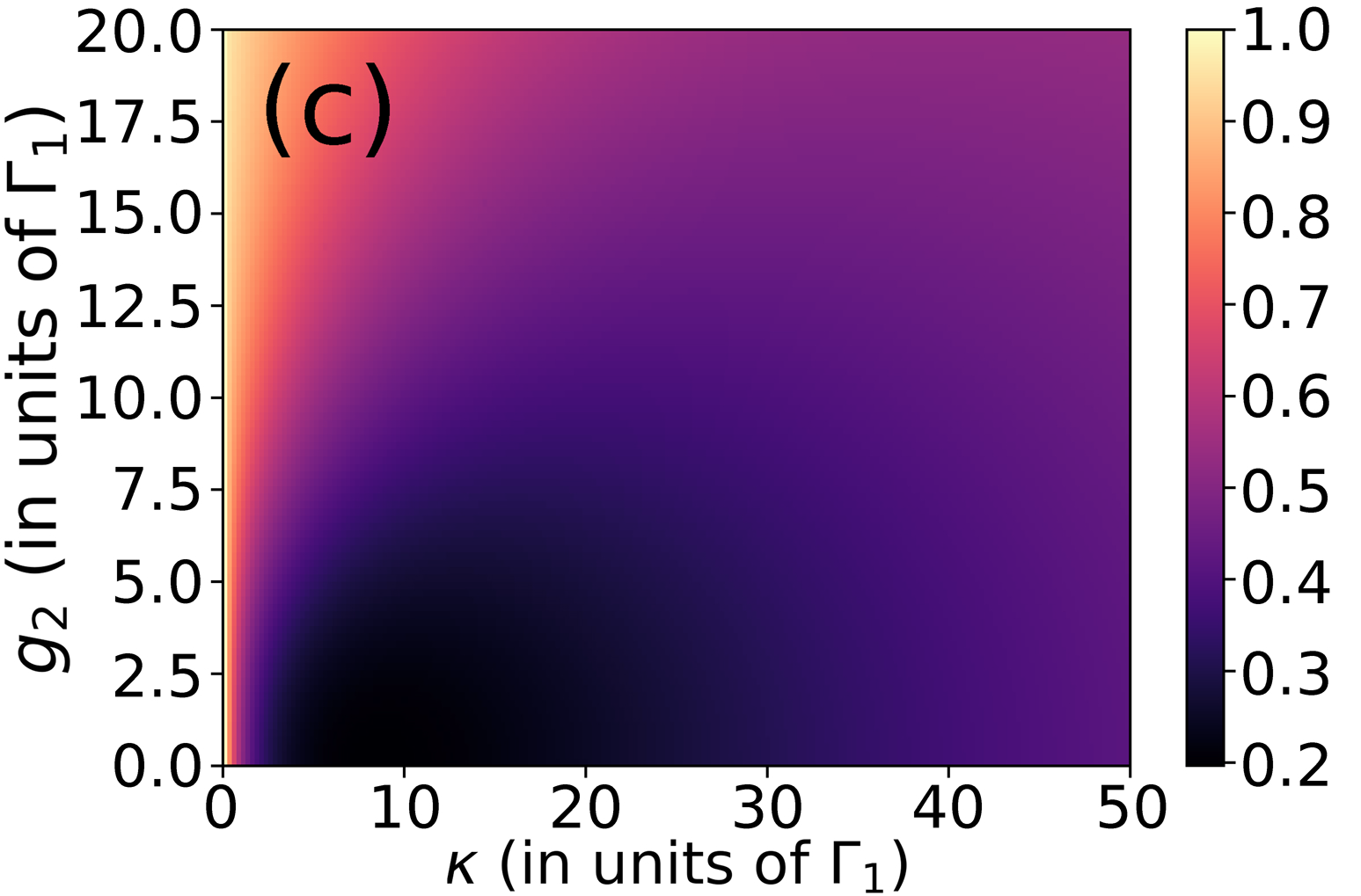}
\label{fig:|eg>+c_decayroute(c)}}
\caption{(color online) Variation of decay probabilities in Eq. (\ref{App_B_em1}-\ref{App_B_emr}) with $\kappa$ and $g_2$. (a) Variation of $P_{\text{em1}}$, $g_1 = g_{12} = \Gamma_1$. (b) Variation of $P_{\text{emx2}}$ = $P_{\text{emr}}$, $g_1 = g_{12} = \Gamma_1$. (c) Variation of $P_{\text{em1}}$, $g_1 = g_{12} = 3\Gamma_1$.}
\label{fig:|eg>+c_decayroute}
\end{figure*}

By symmetry, it can be seen that the effect of qubit 2 and cavity on qubit 1 is equivalent. Thus, we set $g_1 = g_{12}$ and $\Gamma_2 = \kappa$ without loss of generality. The system exhibits three decay channels: atomic dissipation via $\Gamma_1$ and $\Gamma_2$, and cavity dissipation via $\kappa$. The last two are treated to be the same in this section. We first study the control of decay channels by fixing $g_1 = g_{12} = \Gamma_1$, and vary $\Gamma_2$ and $g_2$. The results are plotted in Figs. \ref{fig:|eg>+c_decayroute(a)} and \ref{fig:|eg>+c_decayroute(b)} .

Due to the decoupling effect of $g_2$, an increase in $g_2$ causes qubit 1 to effectively decouple from qubit 2 and cavity. Thus, $P_{\text{em1}}$ increases while $P_{\text{emx2}}$ and $P_{\text{emr}}$ decreases, which is reflected in Fig. \ref{fig:|eg>+c_decayroute} (see Appendix \ref{appendix_eg+c} for detailed expressions of $P_{\text{em1}}$, $P_{\text{emx2}}$ and $P_{\text{emr}}$). For small $g_2$, as $\kappa$ is increased, $P_{\text{em1}}$ decreases to a minimum, and then increase again. The explanation is that as $\kappa = \Gamma_2$ increases, the decay rate of the cavity and qubit 2 increases, causing qubit 1 to have a lowered emission probability. However, when $\kappa = \Gamma_2$ is too large, qubit 1 effectively decouples from its EME, causing $P_{\text{em1}}$ to increase again as it approaches its free space value. This dip in $P_{\text{em1}}$ is not significant for large $g_2$, since qubit 1 is already significantly decoupled due to $g_2$, so the qubit becomes less sensitive to changes in $\kappa$.

We then increase the coupling of qubit 1 by setting $g_1 = g_{12} = 3\Gamma_1$. The results are shown in Fig. \ref{fig:|eg>+c_decayroute(c)}. Since qubit 1 is more strongly coupled to qubit 2 and the cavity, it is expected that its emission probability is more sensitive to the parameters $\kappa$, $\Gamma_2$ and $g_2$. As shown in Fig. \ref{fig:|eg>+c_decayroute(c)}, the dip in $P_{\text{em1}}$ occurs over a bigger range of $\kappa$. Also, the minimum $P_{\text{em1}}$ is lowered from 0.45 in Fig. \ref{fig:|eg>+c_decayroute(a)} to 0.2 in Fig. \ref{fig:|eg>+c_decayroute(c)}. Due to stronger couplings, a larger $g_2$ is required to cause the same effective decoupling, thus the dip is significant even at moderately large $g_2$.

\subsection{Steady state emission spectra}
\label{sect:spectral}

In previous sections, we have studied in detail the control of SE in the temporal domain. If, instead of integrating the probability over all emission frequencies, we take the long-time limit ($t\rightarrow\infty$), the emission spectra can be obtained. Here we report the spectral filter effect where the spectrum of an input photon can be significantly modified by a process involving the spontaneous Raman scattering. This effect may be useful in, for example, converting a broadband input spectrum into an ultranarrow output spectrum. The process in consideration is as follow: a single photon with the input spectrum $\psi(\Delta_2 - \Delta_1)$, where $\Delta_2 = \omega_2 - \omega_{02}$, $\Delta_1 = \omega_1 - \omega_{01}$, is absorbed by qubit 1 to cause the transition $b_1^\dag (\omega) |g_1g_2\rangle \to |e_1g_2\rangle$. The two qubits then exchange excitation $|e_1g_2\rangle \to |g_1e_2\rangle$ via dipole-dipole coupling . Finally, qubit 2 undergoes spontaneous emission, and the resulting photon can be regarded as the Stokes/anti-Stokes photon, depending on the difference between $\omega_{01}$ and $\omega_{02}$. Thus, the final state is $b_2^\dag (\omega_2) |g_1 g_2\rangle$.

Previous work \cite{Muller:2017aa} found that for a $\Lambda$-system with single-photon excitation, the spontaneous Raman spectrum is the product of the the input photon spectrum $\left| \psi(\Delta_2 - \Delta_1) \right|^2$ and the atomic SE spectrum $P_{\text{SE}} (\Delta_2)$. Our calculations further generalise the result in Ref. \cite{Muller:2017aa} by showing that the relation is valid for the cavity coupled two-qubit system by replacing the atomic SE spectrum with the exchange-emission spectrum of qubit 2, $P_{\text{emx2}} (\Delta_2)$. Thus, through suitable tuning of the exchange-emission spectrum of qubit 2 and the relative positions of the qubit and cavity peaks, the output spectrum can be significantly controlled.

We now first study the control of SE in the frequency domain. Following our earlier analysis, $g_2$ primarily functions as an effective decoupling parameter, which we set as zero here without loss of generality. Hence, the area-normalised exchange-emission spectrum for qubit 2 $S_{\text{SE}} (\Delta_2)$ can be written as
\begin{equation}
\label{sp_emx2}
S_{\text{SE}} (\Delta_2) = \frac{1}{\mathcal{N}} P_{\text{emx2}} (\Delta_2),\\
\end{equation}
where $\mathcal{N}$ is the normalisation constant and the probability
$P_{\text{emx2}} (\Delta_2)$ is the long-time limit of $P_{\text{emx2}} (\Delta_2,t)$ in Eq. (\ref{eqn:P_emx2}). In steady-state, the Raman scattering spectrum is
\begin{equation}
\label{Raman}
\begin{split}
S_\text{Raman} (\Delta_2) &= \frac{1}{\mathcal{N}^\prime} g_{12}^2 \Gamma_1 \Gamma_2 \left| \frac{(\omega_2 + i\frac{\kappa}{2} - \omega_c)\psi(\Delta_2 - \Delta_1)}{\prod\limits_{\substack{i = \alpha,\beta,\gamma}} (\omega_2 - \omega_{01} - \omega_{02} - \omega_{i})} \right|^2, \\
&= \frac{1}{\mathcal{N}^\prime} |\psi(\Delta_2 - \Delta_1)|^2 \times P_{\text{emx2}} (\Delta_2)
\end{split}
\end{equation}
where $\omega_{\alpha}, \omega_{\beta}, \omega_{\gamma}$ are the roots of the cubic polynomial below Eq. (\ref{A2}) and $\mathcal{N}^\prime$ is the normalisation constant. For simplicity, we also assume no atomic detuning in the system, i.e. $\omega_{01} = \omega_{02}$.
\begin{figure}
\centering
\subfigure{\includegraphics[width=4cm]{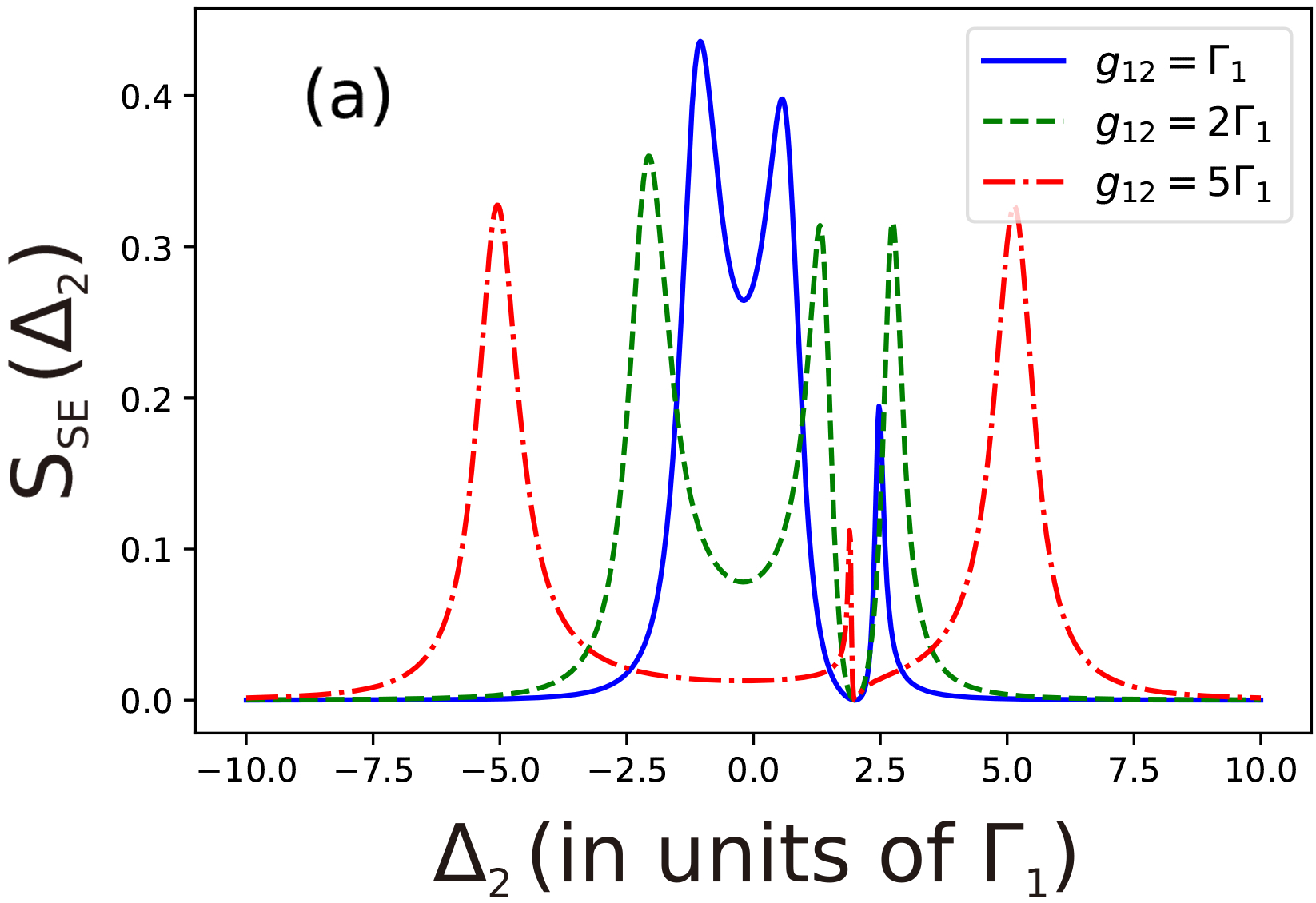}
\label{fig:twoem+c_SE(a)}}
\subfigure{\includegraphics[width=4cm]{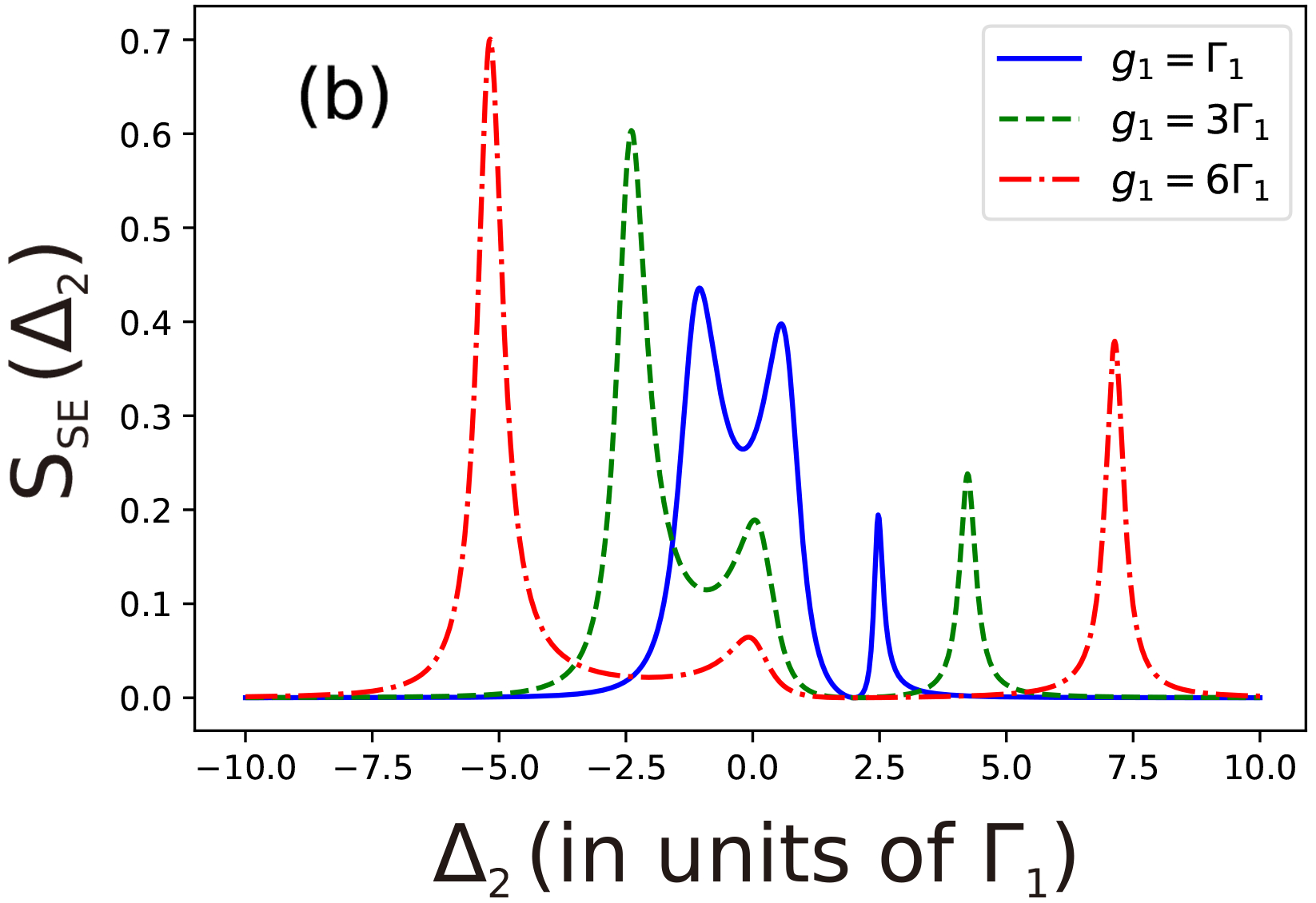}
\label{fig:twoem+c_SE(b)}}
\subfigure{\includegraphics[width=4cm]{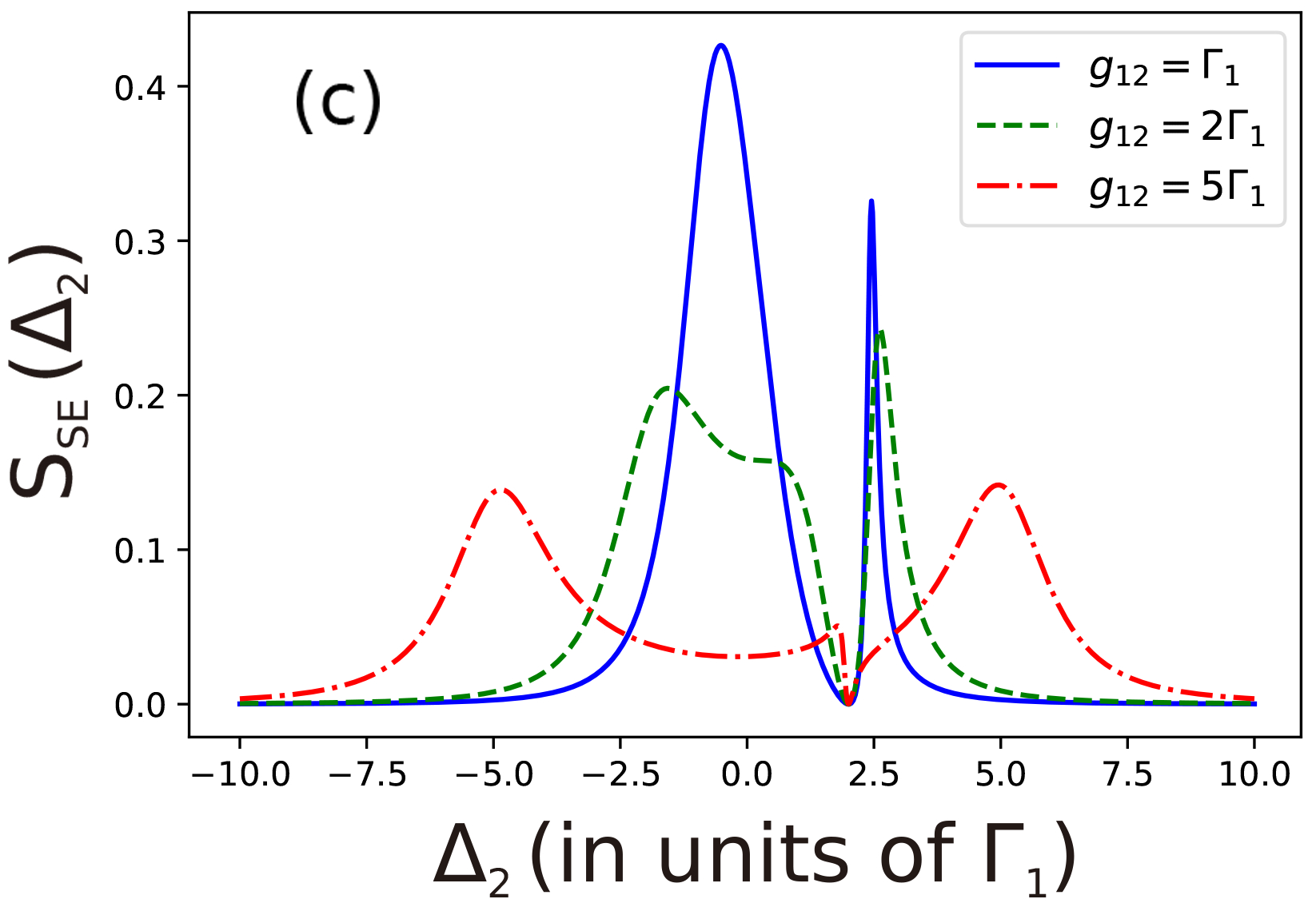}
\label{fig:twoem+c_detuning(a)}}
\subfigure{\includegraphics[width=4cm]{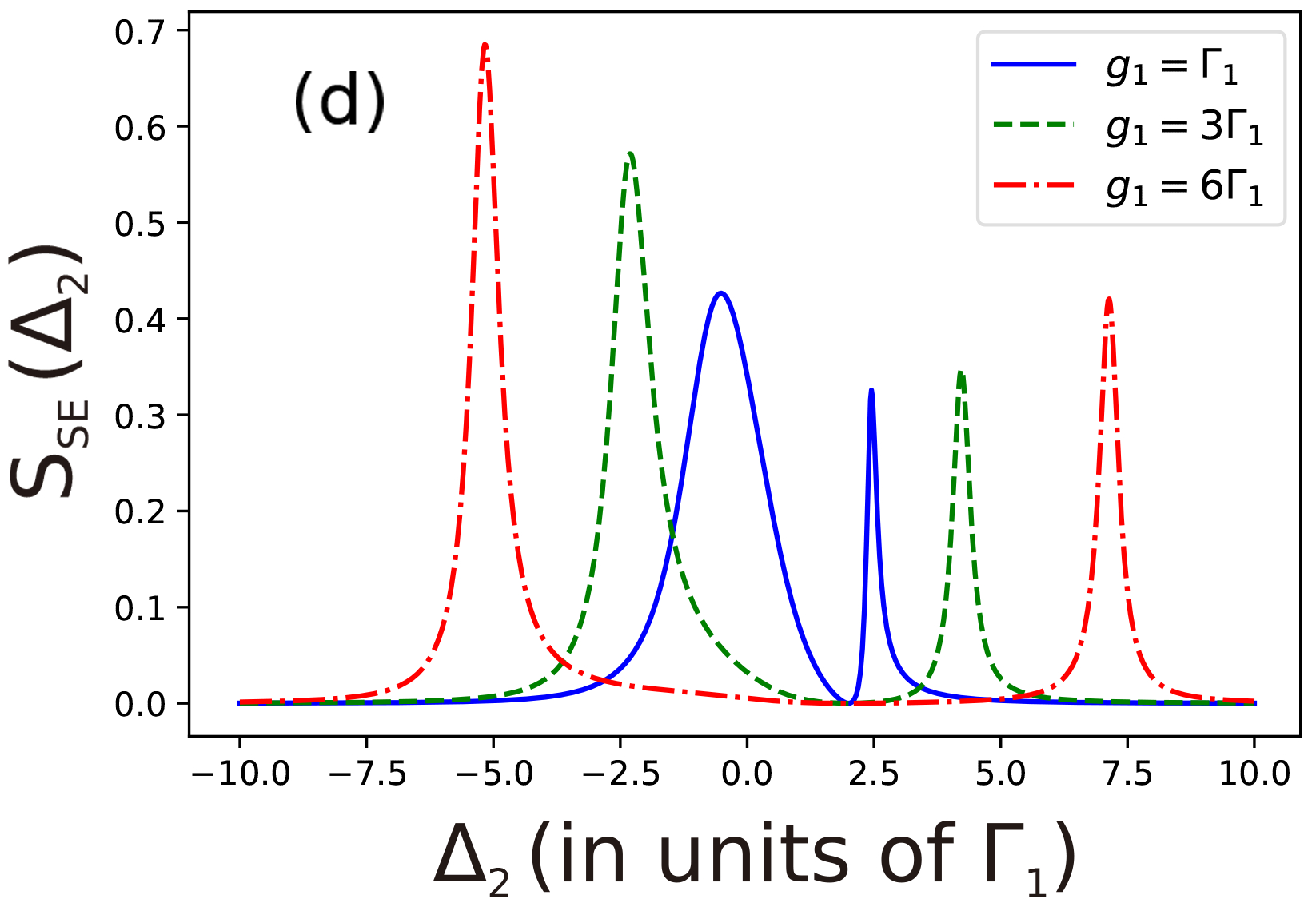}
\label{fig:twoem+c_detuning(b)}}
\caption{(color online) Exchange-emission spectrum in Eq. (\ref{sp_emx2}) of qubit 2. Effect of $g_{12}$ for $\Delta_c = 2\Gamma_1$, $\kappa = 0.01\Gamma_1$, $g_1 = \Gamma_1$ and (a) $\Gamma_2 = \Gamma_1$; (c) $\Gamma_2 = 4\Gamma_1$. Effect of $g_1$ for $\Delta_c = 2\Gamma_1$, $\kappa = 0.01\Gamma_1$, $g_{12} = \Gamma_1$, and (b) $\Gamma_2 = \Gamma_1$; (d) $\Gamma_2 = 4\Gamma_1$.}
\label{fig:twoem+c_SE}
\end{figure}
	
The area-normalised exchange-emission spectrum $S_\text{SE} (\Delta_2)$ contains three peaks. Referring to the blue solid curve in Fig. \ref{fig:twoem+c_SE(a)}, the left peak is associated with qubit 1, the middle peak with qubit 2 and the right peak with cavity. For brevity, we label these peaks as qubit 1 peak, qubit 2 peak, and cavity peak respectively. Thus, increasing $g_{12}$ in Fig. \ref{fig:twoem+c_SE(a)} causes Rabi splitting in the qubit peaks, leaving behind the cavity peak near $\Delta_2 = \Delta_c$. Since $\kappa \ll \Gamma_1$, the cavity peak displays a Fano lineshape. Increasing $g_1$ in Fig. \ref{fig:twoem+c_SE(b)} shifts the qubit 1 peak and cavity peak instead, leaving behind the qubit 2 peak near $\Delta_2 = 0$.

In the case of $\Gamma_2 > (\Gamma_1, g_{12})$, energy in qubit 2 is more likely to dissipate via $\Gamma_2$ than fed back to qubit 1 via $g_{12}$. Fig. \ref{fig:twoem+c_detuning(a)} and \ref{fig:twoem+c_detuning(b)} show the spectrum when $\Gamma_2 = 4\Gamma_1$. In Fig. \ref{fig:twoem+c_detuning(a)}, when $\Gamma_2 \gg g_{12}$, there is only one qubit peak (qubit 1), indicating that qubit 2 has effectively decoupled from the system. The spectrum reduces to that of a JCM. However, when $\Gamma_2 < g_{12}$, the dissipation of qubit 2 does not dominate over the dipole interactions, hence the qubit 2 peak is still observed. In Fig. \ref{fig:twoem+c_detuning(b)}, $g_{12} = \Gamma_1 < \Gamma_2$, thus qubit 2 effective decouples and the spectrum resembles that of a JCM for all values of $g_1$. Therefore, through suitable tuning of parameters, the exchange-emission spectrum can be controlled. In particular, by controlling the Rabi splitting and the relative positions of the qubit and cavity peaks, an ultranarrow Fano peak can be achieved as shown in Fig. \ref{fig:twoem+c_SE(a)} with $g_{12}=5\Gamma_{1}$ which is useful to the spectral filter effect.

As an example, we use a Gaussian input pulse to represent a Fourier-limited wave packet and a pure photonic state \cite{Muller:2017aa}, denoted by $\psi(\Delta_2 - \Delta_1)$. The input spectrum is given by
\begin{equation}
\psi(\omega) = \sqrt[4]{ \frac{2}{\pi \Delta \omega^2} } \text{exp} \left[ - \frac{ (\omega - \omega_1)^2}{\Delta \omega_1^2} e^{i(\omega - \omega_1) \tau} \right]
\end{equation}
 where the pulse duration is $T = 2 \sqrt{3} \Gamma_1^{-1}$, which gives a spectral linewidth $\Delta \omega_1 = \frac{2\sqrt{3}}{T} = \Gamma_1$. The dipole-dipole coupling is set as $g_{12} = \Gamma_1$ and the qubit detuning $\Delta \omega_0 \equiv \omega_{02} - \omega_{01} = 0$ as stated earlier. A small $\kappa$ of $0.05\Gamma_1$ and a cavity detuning of $\Delta_c = \omega_c - \omega_{01} = 2\Gamma_1$ will result in a Fano resonant peak at $\omega_2 = \omega_c$.

\begin{figure}
\centering
\subfigure{\includegraphics[width=4cm]{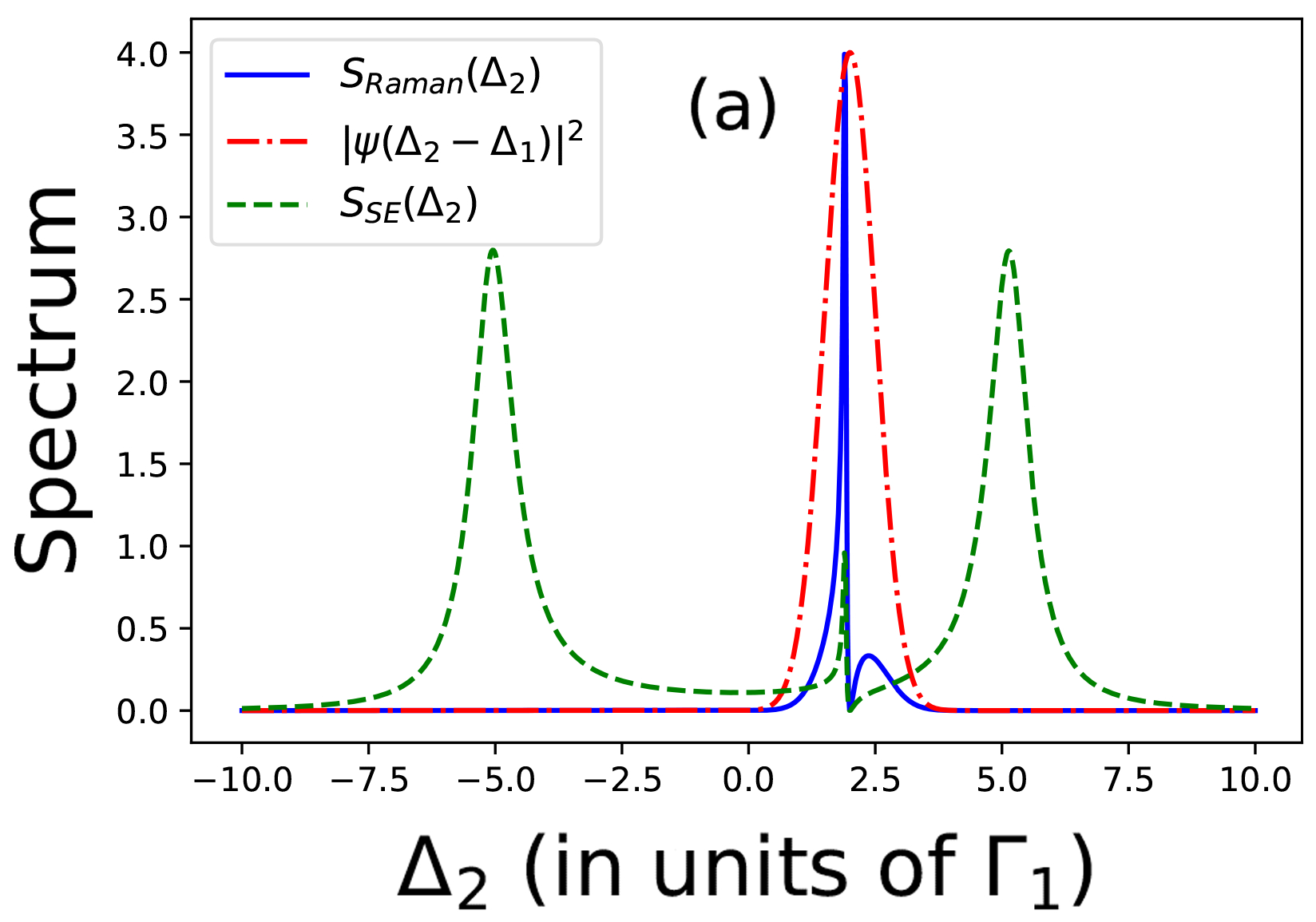}
\label{fig:twoem_raman(a)}}
\subfigure{\includegraphics[width=4cm]{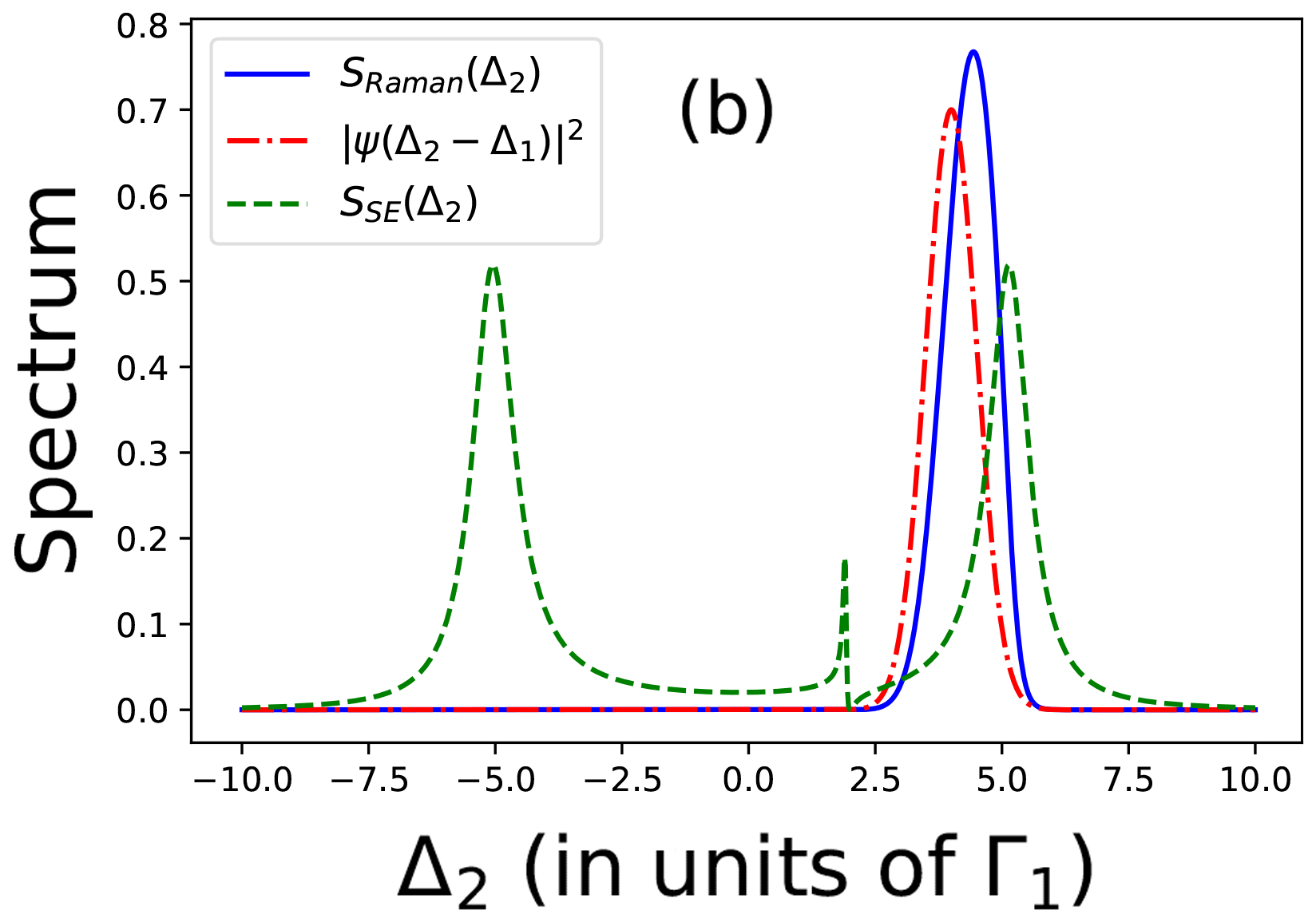}
\label{fig:twoem_raman(b)}}
\caption{(color online) Raman scattering spectrum (blue solid) in Eq. (\ref{Raman}) for $g_1 = \Gamma_1$, $g_{12} = 5\Gamma_1$, $\Delta_c = 2\Gamma_1$, $\kappa = 0.01\Gamma_1$. The red and green curves are the Gaussian and atomic components of the spectrum. Note that the two components are scaled for visualisation purposes. (a) $\Delta_1 = 2\Gamma_1$, (b) $\Delta_1 = 4\Gamma_1$.}
\label{fig:twoem_raman}
\end{figure}

In Fig. \ref{fig:twoem_raman(a)}, when the excitation detuning $\Delta_1 \equiv \omega_1 - \omega_{01} = 2\Gamma_1$, the Gaussian peak (red dash-dotted curve) and Fano peak (green dashed curve) are very near to each other. Hence, the resulting $S_{\text{Raman}} (\Delta_2)$ is similar to the ultranarrow Fano peak. Increasing the excitation detuning to $\Delta_1 = 4\Gamma_1$ shifts the Gaussian peak away from Fano peak. The resulting $S_{\text{Raman}} (\Delta_2)$ is a product of the input Gaussian and atomic Lorentzian as shown in Fig. \ref{fig:twoem_raman(b)} which is not greatly affected by the exchange-emission spectrum of qubit 2. Hence, this system can be used as a spectral filter where the input photon is converted to a Stokes/anti-Stokes photon with a narrow linewidth if the input pulse is tuned appropriately and if the above conditions on $\kappa$ and $\Delta_c$ are met.

Thus, the characterisation and control of the emission peaks of qubit 2 via the system parameters are important in this spectral filter. For example, to utilise the filter effectively and obtain a narrow output linewidth, the two qubit peaks must be sufficiently far apart from the cavity Fano peak. This can be achieved by increasing $g_{12}$ and setting $\kappa \ll \Gamma_1$, as we have done in Fig. \ref{fig:twoem+c_SE}.

\section{Conclusion}
\label{sec:conclusion}

Using the equation-of-motion method, we have explicitly calculated various transition probabilities in the time domain involved in the spontaneous emission of qubit 1. We found that the SE rate of the qubit can be significantly modified by engineering its electromagnetic environments, i.e. surrounding qubit 2 and/or cavity. The Purcell enhancement and inhibition of qubit 1 are both observed in the system. In general, when the decay rate of qubit 1 dominates over the decay rates of surrounding qubit/cavity elements, the SE rate of qubit 1 will be suppressed, or vice versa. Furthermore, a stronger suppression can be obtained by preparing the surrounding EMEs in higher energetic state such that the survival probability is greater than the free-space value with significantly longer duration and suppressed Rabi oscillation. It is exhibited that the best inhibition effect is obtained in the two-qubit system with two excitations in strong coupling regime. The control of decay channel is also studied by explicitly calculating the decay probabilities for each channel. We find that different dominant decay channels of qubit 1 can be realised by fine tuning the parameters of system, i.e., decay rates and coupling rates. This finding may imply some potential applications in quantum devices, such as the photonic switch and router.

In the frequency domain, by studying the Raman scattering spectrum of the TCM, we find that the system can be utilised as a spectral filter. Since the spectrum of the scattered photon is the product of the input pulse spectrum and the atomic response, if the Fano peak (from qubit 2 emission spectrum) is near the input photon peak, the scattering spectrum has a sharp peak around the cavity frequency, thus filtering out most of the non-resonant frequencies from the input spectrum. The Fano peak can be obtained by using large dipole coupling and setting $\kappa \ll \Gamma_1$. The spectral filter effect is not limited to the TCM discussed here and can be easily adapted for other models and input pulse shapes.

In future works, instead of focusing on the two-level system (qubit), we can explore the spontaneous emission in a driven multi-level system where the effects of higher levels \cite{PhysRevA.52.710,PhysRevLett.91.123601,PhysRevLett.81.293} and quantum interference \cite{PhysRevLett.66.2593,Khan2017} are usually important. Our study provides insights to the control and characterisation of spontaneous emission of qubits, which is key to quantum information processing and communication.

\section*{acknowledgments}

J. B. You would like to acknowledge the support by the National Research Foundation Singapore (Grant No. NRF2017NRF-NSFC002-015, NRF2016-NRF-ANR002, NRF-CRP 14-2014-04) and A*STAR SERC (Grant No. A1685b0005).

\bibliography{Bibliography}

\onecolumngrid

\appendix

\section{Emission probabilities for two-qubit system with double excitations}
\label{double_excitation}

In Sec. (\ref{two_qubit_double_excitations}), we have discussed the controllability of decay channels for the SE of the system. There are three sub-processes involved in the decay to the ground state $|{g_{1}g_{2}}\rangle$. The emission probabilities where both photons are emitted from the same qubit, $P_\text{em11}(t, \Delta_1, \Delta_1^\prime) = | \langle{g_1 g_2}| b_1 (\omega_1^\prime) b_1 (\omega_1) U(t) | e_1 e_2 \rangle |^2$ and $P_\text{em22}(t, \Delta_2, \Delta_2^\prime) = | \langle{g_1 g_2}| b_2 (\omega_2^\prime) b_1 (\omega_2) U(t) | e_1 e_2 \rangle |^2$ can be calculated as
	\begin{equation}
\label{em11}
	\begin{split}
P_\text{em11}(t, \Delta_1, \Delta_1^\prime) &= \bigg|g_{12} \frac{\Gamma_1}{2\pi} \bigg[ \frac{1}{(T-\omega_+)(T - \omega_-)(T + i\Gamma)} e^{-iTt} + \frac{1}{(\omega_+ - \omega_-)(\omega_+ - T)(\omega_+ + i\Gamma)} e^{-i\omega_+ t} \\
&+ \frac{1}{(\omega_- - \omega_+)(\omega_- - T)(\omega_- + i\Gamma)} e^{-i \omega_- t} - \frac{1}{(i\Gamma + \omega_+)(i\Gamma + \omega_-)(i\Gamma + T)} e^{-\Gamma t} \bigg] \bigg|^2
	\end{split}
	\end{equation}
	and
	\begin{equation}
\label{em22}
	\begin{split}
P_\text{em22}(t, \Delta_2, \Delta_2^\prime) &= \bigg|g_{12} \frac{\Gamma_2}{2\pi} \bigg[ \frac{1}{(T-\omega_+)(T - \omega_-)(T + i\Gamma)} e^{-iTt} + \frac{1}{(\omega_+ - \omega_-)(\omega_+ - T)(\omega_+ + i\Gamma)} e^{-i\omega_+ t} \\
&+ \frac{1}{(\omega_- - \omega_+)(\omega_- - T)(\omega_- + i\Gamma)} e^{-i \omega_- t} - \frac{1}{(i\Gamma + \omega_+)(i\Gamma + \omega_-)(i\Gamma + T)} e^{-\Gamma t} \bigg] \bigg|^2,
	\end{split}
	\end{equation}
where $\omega_{\pm} = \omega_1 - \frac{1}{2} (\omega_{01} + \omega_{02}) - i\frac{\Gamma}{2} \pm \frac{1}{2} \sqrt{4 g_{12}^2 + (\Delta \omega_0 - i\frac{\Delta \Gamma}{2})^2}$ in $P_\text{em11}$, and $\omega_{\pm} = \omega_2 - \frac{1}{2} (\omega_{01} + \omega_{02}) - i\frac{\Gamma}{2} \pm \frac{1}{2} \sqrt{4 g_{12}^2 + (\Delta \omega_0 - i\frac{\Delta \Gamma}{2})^2}$ in $P_\text{em22}$. Here $T = \omega_1 + \omega_2 - \omega_{01} - \omega_{02}$.
The emission probabilities where each qubit emits one photon, $P_\text{em12}(t, \Delta_1, \Delta_2) = | \langle{g_1 g_2}| b_2 (\omega_2) b_1 (\omega_1) U(t) | e_1 e_2 \rangle |^2$ and $P_\text{em21}(t, \Delta_1, \Delta_2) = | \langle{g_1 g_2}| b_1 (\omega_1) b_2 (\omega_2) U(t) | e_1 e_2 \rangle |^2$ are given by
\begin{equation}
\label{em12}
	\begin{split}
P_\text{em12}(t, \Delta_1, \Delta_2) &= \bigg| \frac{\sqrt{\Gamma_1 \Gamma_2}}{2\pi} \bigg[ \frac{\omega_+ - \omega_1 + \omega_{02} + i\frac{\Gamma_1}{2}}{(\omega_+ - \omega_-)(\omega_+ + i\Gamma)(\omega_+ - T)} e^{-i \omega_+ t} + \frac{\omega_- - \omega_1 + \omega_{02} + i\frac{\Gamma_1}{2}}{(\omega_- -\omega_+)(\omega_- + i\Gamma)(\omega_- - T)} e^{-i\omega_- t} \\
&+ \frac{\omega_1 - \omega_{02} - i\frac{\Gamma_1}{2}+ i\Gamma}{(i\Gamma + \omega_+)(i\Gamma + \omega_-)(i\Gamma + T)} e^{-\Gamma t} + \frac{T - \omega_1 + \omega_{02} + i\frac{\Gamma_1}{2}}{(T - \omega_+)(T - \omega_-)(T + i\Gamma)} e^{-iTt} \bigg] \bigg|^2
	\end{split}
	\end{equation}
	and
	\begin{equation}
\label{em21}
	\begin{split}
P_\text{em21}(t, \Delta_1, \Delta_2) &= \bigg| \frac{\sqrt{\Gamma_1 \Gamma_2}}{2\pi} \bigg[ \frac{\omega_+ - \omega_2 + \omega_{01} + i\frac{\Gamma_2}{2}}{(\omega_+ - \omega_-)(\omega_+ + i\Gamma)(\omega_+ - T)} e^{-i \omega_+ t} + \frac{\omega_- - \omega_2 + \omega_{01} + i\frac{\Gamma_2}{2}}{(\omega_- -\omega_+)(\omega_- + i\Gamma)(\omega_- - T)} e^{-i\omega_- t} \\
&+ \frac{\omega_2 - \omega_{01} - i\frac{\Gamma_2}{2} + i\Gamma}{(i\Gamma + \omega_+)(i\Gamma + \omega_-)(i\Gamma + T)} e^{-\Gamma t} + \frac{T - \omega_2 + \omega_{01} + i\frac{\Gamma_2}{2}}{(T - \omega_+)(T - \omega_-)(T + i\Gamma)} e^{-iTt} \bigg] \bigg|^2,
	\end{split}
	\end{equation}
where $T = \omega_1 + \omega_2 - \omega_{01} - \omega_{02}$, $\omega_\pm = \omega_1 - \frac{1}{2} (\omega_{01} + \omega_{02}) - i\frac{\Gamma}{2} \pm \frac{1}{2} \sqrt{4 g_{12}^2 + (\Delta \omega_0 - i\frac{\Delta \Gamma}{2})^2}$ in $P_\text{em12}$, and $\omega_\pm = \omega_2 - \frac{1}{2} (\omega_{01} + \omega_{02}) - i\frac{\Gamma}{2} \pm \frac{1}{2} \sqrt{4 g_{12}^2 + (\Delta \omega_0 - i\frac{\Delta \Gamma}{2})^2}$ in $P_\text{em21}$.

\section{Emission probabilities for two qubits with cavity, single excitations}
\label{appendix_eg+c}
The survival probability, $P_{\text{surv}} (t) = | \langle{e_1 g_2}| U(t) |{e_1 g_2}\rangle |^2$ is given by
\begin{equation}
\begin{split}
P_{\text{surv}} (t) &= \bigg| \frac{(\omega_\alpha - \omega_+)(\omega_\alpha - \omega_-)}{(\omega_\alpha - \omega_\beta)(\omega_\alpha - \omega_\gamma)} e^{-i \omega_\alpha t} + \frac{(\omega_\beta - \omega_+)(\omega_\beta - \omega_-)}{(\omega_\beta - \omega_\alpha)(\omega_\beta - \omega_\gamma)} e^{-i \omega_\beta t} + \frac{(\omega_\gamma - \omega_+)(\omega_\gamma - \omega_-)}{(\omega_\gamma - \omega_\alpha)(\omega_\gamma - \omega_\beta)} e^{-i \omega_\gamma t} \bigg|^2,
\end{split}
\label{A2}
\end{equation}
where $\omega_\pm = \frac{1}{2} \bigg[ \omega_c - 2\omega_{01} - \omega_{02} - i\frac{\Gamma_2}{2} - i\frac{\kappa}{2} \pm \sqrt{4g_2^2 + (i\frac{\Gamma_2}{2} - i\frac{\kappa}{2} + \omega_c - \omega_{02})^2} \bigg]$ and $\omega_\alpha, \omega_\beta, \omega_\gamma$ are the complex roots of the cubic polynomial of $ip$, $(ip + i\frac{\Gamma_1}{2} + \omega_{02})(ip - \omega_+)(ip - \omega_-) - g_1^2 (ip + i\frac{\Gamma_2}{2} + \omega_{01})$. The emission probabilities $P_{\text{em1}} (t, \Delta_1)$, $P_{\text{emx2}} (t, \Delta_2)$ and $P_{\text{emr}} (t, \Delta_r)$ read
\begin{equation}
\label{App_B_em1}
\begin{split}
P_{\text{em1}} (t, \Delta_1) &= \frac{\Gamma_1}{2\pi} \bigg| \frac{(c_1 - \omega_+)(c_1 - \omega_-)}{(c_1 - \omega_\alpha)(c_1 - \omega_\beta)(c_1 - \omega_\gamma)} e^{-ic_1 t} + \frac{(\omega_\alpha - \omega_+)(\omega_\alpha - \omega_-)}{(\omega_\alpha - c_1)(\omega_\alpha - \omega_\beta)(\omega_\alpha - \omega_\gamma)} e^{-i\omega_\alpha t} \\
&+ \frac{(\omega_\beta - \omega_+)(\omega_\beta - \omega_-)}{(\omega_\beta - c_1)(\omega_\beta - \omega_\alpha)(\omega_\beta - \omega_\gamma)} e^{-i\omega_\beta t} + \frac{(\omega_\gamma - \omega_+)(\omega_\gamma - \omega_-)}{(\omega_\gamma - c_1)(\omega_\gamma - \omega_\alpha)(\omega_\gamma - \omega_\beta)} e^{-i\omega_\gamma t} \bigg|^2,
\end{split}
\end{equation}
\begin{equation}
\begin{split}
P_{\text{emx2}} (t, \Delta_2) &= \frac{\Gamma_2}{2\pi} \bigg| \frac{g_{12} (c_2 + d_2) + g_1 g_2}{(c_2 - \omega_\alpha)(c_2 - \omega_\beta)(c_2 - \omega_\gamma)} e^{-ic_2 t} + \frac{g_{12} (\omega_\alpha + d_2) + g_1 g_2}{(\omega_\alpha - c_2)(\omega_\alpha - \omega_\beta)(\omega_\alpha - \omega_\gamma)} e^{-i\omega_\alpha t} \\
&+ \frac{g_{12} (\omega_\beta + d_2) + g_1 g_2}{(\omega_\beta - c_2)(\omega_\beta - \omega_\alpha)(\omega_\beta - \omega_\gamma)} e^{-i\omega_\beta t} + \frac{g_{12} (\omega_\gamma + d_2) + g_1 g_2}{(\omega_\gamma - c_2)(\omega_\gamma - \omega_\alpha)(\omega_\gamma - \omega_\beta)} e^{-i\omega_\gamma t} \bigg|^2
\end{split}
\label{eqn:P_emx2}
\end{equation}
and
\begin{equation}
\label{App_B_emr}
\begin{split}
P_{\text{emr}} (t, \Delta_r) &= \frac{\kappa}{2\pi} \bigg| \frac{g_1 (c_r + d_r) + g_2 g_{12}}{(c_r - \omega_\alpha)(c_r - \omega_\beta)(c_r - \omega_\gamma)} e^{-ic_r t} + \frac{g_1 (\omega_\alpha + d_r) + g_2 g_{12}}{(\omega_\alpha - c_r)(\omega_\alpha - \omega_\beta)(\omega_\alpha - \omega_\gamma)} e^{-i\omega_\alpha t} \\
&+ \frac{g_1 (\omega_\beta + d_r) + g_2 g_{12}}{(\omega_\beta - c_r)(\omega_\beta - \omega_\alpha)(\omega_\beta - \omega_\gamma)} e^{-i\omega_\beta t} + \frac{g_1 (\omega_\gamma + d_r) + g_2 g_{12}}{(\omega_\gamma - c_r)(\omega_\gamma - \omega_\alpha)(\omega_\gamma - \omega_\beta)} e^{-i\omega_\gamma t} \bigg|^2,	
\end{split}
\end{equation}
where $c_1 = \omega_1 - \omega_{01} - \omega_{02}$, $c_2 = \omega_2 - \omega_{01} - \omega_{02}$, $d_2 = i\frac{\kappa}{2} - \omega_c + \omega_{01} + \omega_{02}$, $c_r = \omega_r - \omega_{01} - \omega_{02}$ and $d_r = i\frac{\Gamma_2}{2} + \omega_{01}$. As before, the emission probabilities are integrated over the respective emission frequencies to obtain the total emission as a function of time.

\end{document}